\documentclass[a4paper,11pt]{article}

\usepackage{graphicx}

\usepackage{bbm}
\def\identity{\mathbbm{1}}

\newcommand{\be}{\begin{equation}}
\newcommand{\eb}{\end{equation}}

\newcommand{\ba}{\begin{eqnarray}}
\newcommand{\ab}{\end{eqnarray}}
\newcommand{\iso}{{\rm iso}}

\def\mnras{Mon. Not. Roy. Astron. Soc.}
\def\apjl{ApJL}

\def\prd{Phys. Rev. {\bf D}}
\def\apj{Astrophys. Journal}
\def\apjs{Astrophys. Journal Supp.}  

\def\vev#1{\left\langle #1\right\rangle} 
\def\brn#1{\left( #1\right)} 
\def\abs#1{\vert #1 \vert}  

\newcommand{\vp}{{\bf \hat p}}
\newcommand{\vpp}{{\bf \hat p'}}
\newcommand{\vk}{{\bf \hat k}}

\def\la{~\mbox{\raisebox{-.6ex}{$\stackrel{<}{\sim}$}}~}

\begin{document}
\begin{flushright}
{arXiv:0707.4179} \\
UMN-TH-2614/07\\
Imperial/TP/07/CRC/02\\
\end{flushright}
\vspace{0.5cm}

\begin{center}
{\LARGE \bf
Inflationary perturbations in anisotropic backgrounds}\\[0.5cm]

{\LARGE \bf
and their imprint on the CMB}\\[1cm]

{
{\large\bf A. Emir G\"umr\"uk\c{c}\"uo\u{g}lu$^{a\,}$\footnote{E-mail:  ahmet@physics.umn.edu}
\,,\,  Carlo R. Contaldi$^{b\,}$\footnote{E-mail:  c.contaldi@imperial.ac.uk}
\,,\, Marco Peloso$^{a\,}$\footnote{E-mail:  peloso@physics.umn.edu}
}
}
\\[7mm]
{\it $^a$ School of Physics and Astronomy,
University of Minnesota, Minneapolis, MN 55455, USA
}\\[3mm]
{\it $^b$ 
Theoretical Physics, Blackett Laboratory, Imperial College, London, SW7 2BZ, UK 
}
\\[1cm]
\vspace{-0.3cm}

\vspace{1cm}

{\large\bf Abstract}

\end{center}
\begin{quote}

{We extend the standard theory of cosmological perturbations to
homogeneous but anisotropic universes. We
present an exhaustive computation for the case of a Bianchi $I$ model,
with a residual isotropy between two spatial dimensions,  which is 
undergoing complete isotropization at the onset of inflation; we also
show how the computation can be further extended to more general
backgrounds. In presence of a single inflaton field, there are three physical
perturbations (precisely as in the isotropic case), which are obtained
(i) by removing gauge and nondynamical degrees of freedom, and (ii) by
finding the combinations of the remaining modes in terms of which the
quadratic action of the perturbations is canonical. The three
perturbations, which later in the isotropic regime become a scalar
mode and two tensor polarizations (gravitational wave),
are coupled to each other already
at the linearized level during the anisotropic phase. This generates
nonvanishing correlations between different modes of the CMB
anisotropies, $\langle a^{\,}_{\ell m}a^\star_{\ell' m'}\rangle
\propto \!\!\!\!\! \not \;\;\;\; \delta_{\ell\ell'}\delta_{mm'}$,
which can be particularly relevant at large scales (and, potentially,
be related to the large scale anomalies in the WMAP data).  As an
example, we compute the spectrum of the perturbations in this 
Bianchi $I$ geometry, assuming that the inflaton is in a 
slow roll regime also in the
anisotropic phase. For this simple set-up, fixing the initial
conditions for the perturbations appears more difficult than in the
standard case, and additional assumptions seem to be needed to provide
predictions for the CMB anisotropies.  }

\end{quote}


\section{Introduction}\label{sec:intro}

The study of Cosmic Microwave Background (CMB) full sky maps from the
WMAP experiment \cite{spergel06} has led to some intriguing anomalies
which seem to suggest that the assumption of statistical isotropy is
broken on the largest angular scales \cite{anomalies}. These anomalies
include an alignment of the moments in the lowest multipoles dubbed the `axis
of evil', an asymmetry in the power between the northern and southern
ecliptic hemispheres and an apparently non-Gaussian excursion in the
southern galactic hemisphere known as the `cold-spot'. Another
well-known puzzle in the observations has been the lack of power in
the quadrupole which had also been noted in the COBE-DMR
maps \cite{cobe} and is still present in the latest WMAP data 
\cite{wmap3}.

The significance of the anomalies has been debated extensively in the
literature (see e.g. \cite{significance}) with some reported effects
more significant than others. The difficulty in quantifying exactly
the importance of any effect is due to how correctly the {\sl a
posteriori} probability of observing it is estimated. It is clear
however that the anomalies implying an overall anisotropy in the data
are more significant than the lack of power in the CMB quadrupole.  A
natural explanation for the observed anomalies may be some form of as
yet undetermined systematic or foreground signal which is not being
taken into account properly in the data reduction producing the final
maps \cite{foregrounds}. However, a conclusive explanation along these
lines has not been put forward yet. It is therefore legitimate to ask
whether the observed anomalies may be an indication of a departure
from the standard cosmological model. In order to find convincing
evidence on such deviations, one should study the CMB properties
predicted by any specific model, with a particular emphasis on those
which go beyond the ones characterizing the standard picture. For
instance, if the signal {\sl is} statistically anisotropic, crucial
information will be encoded in the off-diagonal correlations of the
spherical harmonic modes with $\langle a^{\,}_{\ell m}a^\star_{\ell'
m'}\rangle \propto \!\!\!\!\! \not \;\;\;\;
\delta_{\ell\ell'}\delta_{mm'}$, which are predicted to vanish in the
standard case.

Since, aside the anomalies \cite{anomalies}, CMB data 
strongly support   
the general theory of inflation, we focus here on departures
from the standard picture which can however
be reconciled with the inflationary framework. For example,
the low quadrupole power has motivated a number of studies to explain
the observed cut-off within the context of inflation
\cite{contaldi03,Cline03,Piao04,Kinney06} 
(ref.~\cite{donoghue} further extended the model~\cite{contaldi03}, 
to account for a spatial
asymmetry of the CMB perturbations at large scales). 
These models assume a short
period of inflation, leading to a fine tuning problem in the standard
inflationary picture since the slow-roll regime responsible for the
observed $N\sim 60$ or so $e$-folds is an attractor in the inflaton
phase space and thus the probability of inflating for a large number
of $e$-folds is much larger than that of inflating for only 60.  
\footnote{See however other inflation models that
naturally suppress the overall number of $e$-folds, e.g.
\cite{Kaloper04}.} Recently, other groups have argued that inflation
should be short-lived \cite{Gibbons06} and in particular models of
inflation on the string theory landscape seem to predict a highly
suppressed probability of large $N$ \cite{freivogel06}. Coupled to the
fact that our observed universe requires $N\approx 60$, these arguments
imply that inflation must have lasted `just long enough'. 
Theoretical arguments aside, a limited amount of inflation would certainly
lead to a richer phenomenology than a prolonged one. The reason why inflation
was originally postulated is that it leads to the isotropic and homogeneous
Friedmann-Robertson-Walker (FRW) universe starting from rather generic initial conditions.
If inflation lasted many $e$-folds, then any trace of the pre-existing universe
is inflated away to scales much larger than the present day
horizon. On the other hand, a limited amount of inflation could have left
some trace in the data, particularly at large scales,
 that would appear as anomalies within the standard
inflationary picture. 

With these considerations in mind, in this work we perform an initial
step towards the study of the CMB anisotropies taking into account a
pre-existing phase before inflation, in which the universe is not yet
of the FRW type. Motivated by the suggested alignment of the large scale multipoles
along a common axis, we focus on models which are anisotropic at the
onset of inflation. For simplicity, we still assume that the universe
is homogeneous. As we will see, even this very minor departure from
standard cosmology requires a significant extension of the well
consolidated framework for the computation of CMB anisotropies in a
FRW universe.  Homogeneous but anisotropic cosmologies have long ago
been classified into equivalence classes known as the Bianchi types
\cite{bianchi}.  With the possible exception of type $IX$, Bianchi
universes with a positive cosmological constant evolve towards an
asymptotic (isotropic) de Sitter stage~\cite{wald}. The situation is
more complicated when the cosmological constant is replaced by an
inflaton field (which one has to do, in order to have a finite amount
of inflation). Also in this case, however, one finds that inflation
takes place for a wide variety of circumstances, leading to the
isotropization of the space.~\footnote{ A list of works discussing
this issue, and several aspects of inflation in Bianchi models can be
found in the review~\cite{keith}.} 

It was shown in~\cite{bianchi7} that a late time contribution of the Bianchi $VII_h$
form to the temperature anisotropies can account for several of the WMAP anomalies;
more in general, Bianchi models with
generalized backgrounds driven by anisotropic stresses have been
studied extensively in the literature (see
e.g. \cite{barrow_anis,Berera}). These models require a particular
mechanism actively driving the anisotropy, which can 
for this reason survive for a much longer period than in the case considered here
(and, possibly, result in stronger observational effects). In this work we consider
only the case consistent with the simplest models of inflation where
the background is initially anisotropic and is isotropized at the onset 
of inflation.~\footnote{We also assume that the only source is the inflaton field.
Ref. \cite{gianpiero} studied the background evolutions in cases in which an anti-symmetric
tensor is also dynamically relevant.} We expect to recover a standard 
power spectrum at the scales that leave the horizon after the isotropization
has been achieved, but nonconventional results at larger scales.~\footnote{Ref.~\cite{wise} 
studied the perturbations of a scalar field assuming that a small anisotropy 
(corresponding to the ratio $h/H$ in our equation~(\ref{defHh})) is present during inflation, 
and that it then decays away at the end of inflation. Such perturbations are then converted into
metric perturbations through the mechanism of modulated 
perturbations~\cite{lev}. This work presents analytical results for the perturbations
using an expansion series in the anisotropy parameter, which, in our context,
is valid for modes leaving the horizon towards the
end of the anisotropic phase.}

Thus, the anisotropy is restricted to the initial conditions, in the
form of a power spectrum of the primordial perturbations which depends
also on the directionality of the modes, and not only on the magnitude
of their momentum ${\bf k}$ (as in the isotropic case). On the
contrary, the transfer functions, relating the late time temperature
anisotropies to the primordial perturbations, are the standard ones,
since the propagation of the modes, soon after the onset of inflation,
occurs in an isotropic background. This simplifies greatly the
calculation of the correlations induced in the CMB.~\footnote{This is
in contrast to the case where the anisotropy is present during the
evolution of perturbations at any time in the radiation through to the
present epoch. In this case the Einstein-Boltzmann system must be
modified to account for the anisotropic evolution of the modes
\cite{bunn96,gordon05,Pontzen:2007ii}.} We present this computation in
Section~\ref{sec-cllmm}, in the case in which there is a residual
anisotropy between two of the three spatial directions.

Namely, we compute the correlation between different CMB modes, under
the assumption that (i) the primordial perturbations are Gaussian, and
(ii) their power spectrum depends on the magnitude $\vert {\bf k}
\vert$ and on the angle between the momentum and the anisotropic
direction. We do not impose any further assumption in this
computation, so that it holds for any Bianchi model with a residual
$2$d isotropy and an asymptotic standard inflationary behaviour. As we
anticipated a few paragraphs above, the main result is a non vanishing
correlation between off diagonal modes, see eq.~(\ref{corrcllmm}). The
specific relations found between the different non vanishing
correlators can in principle allow for a detailed data analysis within
any given model. We do not perform this analysis in the present work (since, as
we mentioned, our main result is the extension of the computation of
the primordial perturbations to such non FRW geometries); however we
include the relations~(\ref{corrcllmm}), since they offer a model
oriented way of extracting information from the CMB data. This
information is lost in the standard diagonal $C_\ell$ statistic
analysis. The observed correlations can be fit with model-independent
templates (see e.g. \cite{template,spergel06}) but these are not
calculated {\it a priori} from any particular theory and thus the
information contained in the CMB will not constrain any fundamental
parameter.

In the remainder of the work, we present the detailed computation of
the primordial perturbations  within a Bianchi $I$ model
with a residual $2$d isotropy, which is the
simplest deviation from the FRW geometry. The line element is of the
form
\be
ds^2 = - dt^2 + a \brn{ t }^2 dx^2 + b \brn{ t }^2 \brn{ dy^2 + dz^2 }\,,
\label{background}
\eb
and late time isotropy ($a = b$) is achieved due to an inflaton scalar field $\phi \,$. In Section~\ref{sec-bck} we study the background evolution for this model, under the assumption that the inflaton is initially in a slow roll regime. As it is well known, the anisotropy is rapidly damped away, on a timescale corresponding to the inverse Hubble rate due to the potential energy of $\phi \,$. For this reason, in order to have a sizeable effect, we need to start from a significant anisotropy. The model then admits two possible background solutions at early times, according to whether the expansion rate of the scale factor $a$ is greater or smaller than the one of the scale factor $b$. In the first case (which we call positive branch), $a \propto t$ at asymptotically early times, while $b$ is constant. In the second case (which we call negative branch), the anisotropic direction is initially contracting, $a \propto t^{-1/3} \,$, while the other two directions expand as $b \propto t^{2/3} \,$. In both cases, the metric is singular at $t =0 \,$, and the early time evolution is of the Kasner~\cite{kasner} type.
 
In Section~\ref{pert} we compute the primordial perturbations about this geometry, by extending to this case the Mukhanov-Sasaki~\cite{musa} linearized computation and quantization procedure for
the perturbations valid for a FRW background. 
~\footnote{Perturbations in Bianchi models have been previously studied in~\cite{tode} 
(through the formalism of~\cite{bardeen}) and in~\cite{dunsby} (through the formalism of~\cite{ellisbruni}). These analyses do not compute the canonical variables of the actions of the perturbations,
as instead done in~\cite{musa} for the FRW case, which, as we shall see, is a necessary step to compute the initial conditions for these modes.} For clarity, we summarize at various stages the results of the isotropic computation~\cite{mfb}, and we then show how each of them extends to our case. We start by showing that also in the present case there are three physical modes of the perturbations. We see this with a simple counting of modes, which holds in the standard case, but which does not rely on the isotropy (nor homogeneity) of the background. For a FRW geometry, the three modes are encoded in a scalar perturbation, and in the two polarizations of a tensor mode (gravitational wave). These modes are decoupled at the linearized level, due to the homogeneity and isotropy of the background. In our case, only one mode is decoupled, due to the residual $2$d isotropy, while the two remaining ones are coupled to each other. As the background becomes isotropic, the latter two become the scalar mode, and one of the two tensor polarizations. To extract the physical modes, we choose a non-conventional gauge that preserves all the $g_{0 \mu}$ metric perturbations. Such modes are nondynamical, and can be readily integrated out. Since there are initially $11$ perturbations (symmetric $\delta g_{\mu \nu}$ and $\delta \phi$), we are left with the $11 - 4 ({\rm gauge}) - 4 ({\rm nondynamical}) = 3$ physical modes. Although we perform the explicit computation only for the Bianchi $I$ model with a residual $2$d isotropy, we emphasize that this procedure can be further extended to more general geometries. In the general case, one expects that all the three physical modes are coupled to each other at the linearized level.

To compute the power spectrum of these modes, we need to study their early times frequency. In the standard case, the physical frequency is initially dominated by the physical momentum, and it changes only very slowly (adiabatically) due to the expansion of the universe. In conformal time $\eta$, related to the physical time   $t$ by $d t = a \, d \eta$ the (conformal) frequency is actually constant at early times. Therefore, one can consistently start from an adiabatic vacuum in the asymptotic past (this, in turn, leads to a nearly scale invariant primordial power spectrum). The situation is more complicated for the two Bianchi $I$ backgrounds that we have described above; first of all there is some ambiguity in the choice of conformal time (since one may use different powers of the two scale factors in the definition). This has lead us to study the condition under which two different conformal times can be employed, and result in the same prescription for the initial conditions. We find that, in the negative branch, the frequencies of two coupled modes are not adiabatically evolving at early times; it is possible to find a time variable in which the frequency of the decoupled mode is adiabatically evolving; however, the mode is tachyonic, leading to an instability and to a nonlinear regime that prevents us from computing firm predictions for the CMB anisotropies in it. For the positive branch, ($a \propto t ,\, b \rightarrow {\rm const.}$), all the three modes are decoupled at asymptotically early times, and their three frequencies are adiabatically evolving in the conformal time variable $\eta$, defined as $d t = a \, b^\gamma \, d \eta$ (where $\gamma$ any constant; for  simplicity we then choose $\gamma = 0$ in the computation). This allows us to consistently start from an adiabatic vacuum. The power spectrum that we obtained approaches isotropy for the modes which exit the horizon right after the universe becomes isotropic. At larger scales, however, the spectrum exhibits an angular dependence, and it actually diverges for momenta that are aligned along the directions which are not expanding at asymptotically early times. As a consequence, a computation of the temperature anisotropies requires that we resolve this singularity. In the concluding Section~\ref{conc}, we discuss several ways in which this can be done (at the very least, this will occur due to nonlinear effects, since the singularity of the modes indicates a breakdown of the linearized computation). However, all of them require additional input in the computation. We therefore conclude that the Bianchi $I$ model we have studied is too simple to provide firm predictions for the anisotropy, at least in the assumption of initial slow roll of the inflaton that we have made in this analysis.

The paper is concluded by some Appendixes, where we outline several intermediate steps of the computations.


\section{The $a_{\ell m}$ covariance} \label{sec-cllmm}

In this Section we compute the statistical correlation between different 
multipoles of the CMB temperature anisotropies, relaxing the usual assumption
of statistical isotropy. The starting point is the power spectrum
\begin{equation}
  \vev{{\cal R}^{\,}({\bf k}) {\cal R}^\star({\bf k'})} \equiv
  \frac{(2\pi)^3}{k^3}\delta({\bf k}-{\bf k'})P({\bf k})
\label{power}
\end{equation}
of the primordial comoving curvature perturbations (we assume that the
perturbations are Gaussian). If the background is isotropic, the
fluctuations are statistically isotropic, and the power spectrum $P$
depends only on the magnitude of ${\bf k} \,$. In general, it will depend
also on its direction.

In the following Sections, we restrict our attention to a Bianchi $I$
cosmology. The present computation is instead valid for more general
anisotropic backgrounds, with the assumptions that (i) there is a
residual isotropy of two of the three spatial dimensions, and (ii)
this anisotropy is damped away during inflation. It is straightforward
to generalize the calculation if condition (i) is relaxed. Under these
assumptions, the primordial power spectrum depends on the magnitude of
${\bf k} \,$, and on the angle between ${\bf k}$ and the anisotropic
direction. We denote by $\xi$ the cosine of this angle.

We denote by $\delta T(\vp,\eta_0,{\bf x}_0)$ the temperature
perturbation in the direction $\vp$, as measured by an observer at
position ${\bf x}_0$ and at conformal time $\eta_0 \,$. It is customary to
decompose the temperature perturbations that we observe
into spherical harmonics:
\begin{equation}
\delta T(\vp,\eta_0,{\bf x}_0) = \sum_{l,m} a_{lm} \, Y_{\ell m} \brn{\vp}
\label{dt}
\end{equation}
Since we are assuming that the perturbations are Gaussian, their
statistical properties are still encoded in the second order 
correlations
\begin{equation}
C_{\ell\ell' m m'} \equiv   \vev{a^{\,}_{\ell m}a^\star_{\ell'm'}}.
\label{cllmm}
\end{equation}
The correlations however are not diagonal as in the isotropic case.

In the linear regime, the temperature anisotropy~(\ref{dt}) is given by 
\begin{equation}
\delta T(\vp,\eta_0,{\bf x}_0) = \int \frac{d^3{\bf k}}{(2\pi)^3} {\cal R} ({\bf
  k},\eta_i)\Delta(k,\vk\cdot\vp,\eta_0)  e^{i{\bf k}\cdot{\bf x}_0},
\label{dt2}
\end{equation}
where the transfer function $\Delta(k,\vk\cdot\vp)$ describes the
change in amplitude of the radiation perturbation from an initial time
$\eta_i \,$, which can be taken deep in the radiation dominated era, to today.
Since the background is isotropic soon after the onset of inflation, the transfer
function is also isotropic (in real space). As, a consequence $\Delta$ depends only
on the magnitude of the wavenumber $k = \abs{\bf k}$ of the mode, and on the angle between
${\bf k}$ and the line of sight at which the photon is being observed. 
It is convenient to expand it in a basis of Legendre polynomials: 
\begin{equation}
\Delta(k,\vk\cdot\vp,\eta_0) = \sum_\ell (-i)^\ell (2\ell+1)
P_\ell(\vk\cdot\vp) \Delta_\ell(k,\eta_0).
\label{deltadec}
\end{equation}

We rotate our coordinate system such that the anisotropic direction
lies on the $z-$axis. Starting from the above expressions, after 
the algebra outlined in appendix \ref{appA}, we find
\begin{eqnarray}
C_{\ell \ell' m m'} &=& \frac{\delta_{m m'}}{\pi} \left( - i \right)^{\ell - \ell'} \sqrt{\frac{(2\ell+1)(2\ell'+1)(\ell-m)!(\ell'-m)!}{(\ell+m)!(\ell'+m)!}} \nonumber\\
&& \;\;\;\;\;\;\;\;\; \times \int \frac{d k}{k} \Delta_\ell \left( k , \eta_0 \right) \Delta_{\ell'} \left( k , \eta_0 \right) \int_{-1}^1 d \xi P_\ell^m \left( \xi \right) P_{\ell'}^m \left( \xi \right) P \left( k ,\, \xi \right),
\label{corrcllmm}
\end{eqnarray}
where the $P_\ell^m$ are associated Legendre polynomials. The above
expression conveys all the statistical information present in the CMB
in the axisymmetric anisotropic case. When constraining anisotropic
models the full covariance should be compared to the data.

Some general properties of the correlations follow from Eq.~\ref{corrcllmm}. 

(i) The correlations reduce to the standard result in the isotropic case, $P
\left( k ,\, \xi \right) \equiv P \left( k \right) \,$
\begin{equation}
C_{\ell \ell' m m'}^{({\rm iso})} = \frac{2}{\pi} \delta_{\ell \ell'} \delta_{m m'} \int \frac{d k}{k} P \left( k \right) \Delta_\ell \left( k ,\, \eta_0 \right)^2
\label{cllmmiso}
\end{equation}
in which different multipoles are uncorrelated.  In any realistic model, the power spectrum becomes isotropic at small scales (large $k$), since such modes exit the horizon when the background is isotropic . Small scales correspond to large $l$ (mathematically, this is due to the transfer functions, which are peaked at increasingly large $k$ as $l$ increases). Therefore, off-diagonal correlators become progressively smaller as $l$ and $l'$ increase, and the result~(\ref{corrcllmm}) reduces to the isotropic one. The scale at which this happens is strongly sensitive to the amount of inflation which takes place after the universe has become isotropic. If this stage is too prolonged, the power spectrum is anisotropic only at too large scales to be observed today, and the isotropic result~(\ref{cllmmiso}) is recovered for all multipoles.

(ii) The residual $2$d isotropy of the background restricts the number
of non vanishing correlators. The restriction is mostly manifest in the
coordinate system considered here, where the anisotropic direction coincides with
the $z-$axis, and the multipoles are uncorrelated unless $m = m'
\,$. For a general orientation of the anisotropic direction we find a
greater number of non vanishing correlators. However (obviously) the
number of linearly independent correlators remains the same.

(iii) Bianchi $I$ cosmologies have planar reflection symmetries
(${\bf x} \rightarrow - {\bf x}$), and, as a consequence, $P ( k, - \xi) = P
( k, \xi)$. In this case, the correlation (\ref{corrcllmm}) also
vanishes whenever the difference between $l$ and $l'$ is odd
(mathematically, this follows from the parity properties of the
associated Legendre polynomials).

Naturally, the orientation of the anisotropic direction is not expected to be
correlated with any local preferred direction; for any coordinate system,
the orientation can be given in terms of the  Euler angles $(\alpha,\beta,\gamma)$; 
these are extra free parameters in anisotropic models, which need to be fit to the
data in addition to the usual physical parameters ${\cal P}$
determining the radiation transfer function and the spectral
parameters ${\cal S}$ determining the primordial power spectrum.

In principle, the anisotropic models can be fit to the observed
correlations in the data by evaluating a likelihood of the form 
\begin{equation}\label{eq:likeli}
  L\left[{\bf a}|{\bf C}({\cal P},{\cal S},\alpha,\beta,\gamma)\right] = \frac{1}{\sqrt{(2\pi)^N |{\bf C}|}} \exp\left(
  -\frac{1}{2} {\bf a}^{\dagger} \cdot {\bf C}^{-1}\cdot{\bf a} \right),
\end{equation}
where $N$ is the number of modes in the expansion, ${\bf a}$ is the
vector of observed spherical harmonic coefficients and ${\bf C}$ is
the model correlation (Eq.~\ref{corrcllmm}) between $\ell, m$ and
$\ell',m'$ pairs. The orientation of the anisotropic direction can be
rotated using the spherical harmonic rotation operators ${\cal
D}^\ell_{mm'}(\alpha,\beta,\gamma)$ \cite{Varshalovich}. For the
axisymmetric case one of the Euler angles is degenerate and can be
marginalized out; this is not the case for general anisotropy.
  
A full maximum likelihood search of (\ref{eq:likeli}) is
computationally intensive since the dimensionality of the matrices
involved scale as $(\ell +1)^2$. In addition, cut-sky effects
complicate the rotation of the data vector which is computationally
much faster than multiple rotations of the model correlations ${\bf
C}$. For this reason we postpone any detailed analysis of the
likelihood surface and focus here on the framework required to compute
the correlations in the axisymmetric anisotropic case. As we will see
below, this requires a careful treatment of both the theory of
perturbation evolution in an anisotropic background and the early-time
initial conditions of the perturbations.


\section{The Background} \label{sec-bck}

Here and in the remainder of the paper we focus on a Bianchi $I$ model, with equal expansion rate in two of the three spatial dimensions:
\be
ds^2 = - dt^2 + a \brn{ t }^2 dx^2 + b \brn{ t }^2 \brn{ dy^2 + dz^2 }
\label{bianchids}
\eb
(contrary to the previous Section, we take the anisotropic direction to coincide with the $x$-axis).
This model describes an anisotropic universe as long as the ratio between
the two different scale factors evolves with time (since any constant ratio can be rescaled to one). The two physical quantities characterizing the geometry are the two expansion rates
\begin{equation}
H_a \equiv \frac{\dot{a}}{a} \;\;,\;\; H_b \equiv \frac{\dot{b}}{b}
\label{defhahb}
\end{equation}
where, as usual, dot denotes derivative with respect to physical time $t \,$.

Since we want to study inflation in this background, we consider a scalar
field $\varphi$, with the potential $V$. The action of the system is
\begin{equation}
S = \frac{M_p^2}{2} \int d^4 x \sqrt{-g} \, R + \int d^4 x \sqrt{- g} \left[ - \frac{1}{2} g^{\mu \nu} \partial_\mu \varphi \, \partial_\nu \varphi - V \right]
\label{action}
\end{equation}
We decompose the scalar field as $\varphi = \phi + \delta \phi$, namely in a background quantity $\phi$ (that we take to be homogeneous, as the underlying geometry), plus perturbations $\delta \phi \,$. In the present Section we discuss the background evolution, so we only include $\phi $ here. The scalar field perturbations will be considered in the remainder of the paper, when we also perturb the background metric~(\ref{bianchids}).

The background evolves according to the three independent equations~\footnote{The first 
of~(\ref{bg.system}) is obtained by combining the three nontrivial Einstein equations (namely, $3 {\rm Eq.}^0_0 - {\rm Eq.}^1_1 - 2 {\rm Eq.}^2_2$, where ${\rm Eq.}^i_j$ refers to the $ij$ Einstein equation), while the third one is the $00$ Einstein equation. The second one is the equation for the scalar field (any other linear combination of ${\rm Eq.}^0_0$, ${\rm Eq.}^1_1$, and ${\rm Eq.}^2_2$ can be obtained from~(\ref{bg.system}), as a consequence of the $0$ component Bianchi identity).}
\ba
\dot{H} + 3 \, H^2 & = & V / M_p^2 \,,
\nonumber\\
\ddot{\phi} + 3 \, H \, \dot{\phi} & + & V^\prime = 0 \,,
\nonumber\\
3 \, H^2 - h^2 & = & \frac{1}{M_p^2} \left[ \frac{1}{2} \, \dot{\phi}^2 + V \right] \,,
\label{bg.system}
\ab
where
\begin{equation}
H \equiv \frac{H_a + 2 \, H_b}{3} \;\;\;,\;\;\;
h \equiv \frac{H_a - H_b}{\sqrt{3}}
\label{defHh}
\end{equation}
 ($M_p$ is the reduced Planck mass, while a prime 
on the potential 
denotes differentiation with respect to $\phi$). The first two
equations are identical to the ones obtained in the isotropic case, in
terms of the ``average'' expansion rate $H \,$. The third equation,
appearing as a ``modified Friedmann equation'', can be then used as an
algebraic equation for the difference $h$ between the two expansion
rates. By differentiating
this equation, and by combining it with the other two, one can also find
\be
h \brn{ \dot{h} + 3 \, H \, h} =0 \,,
\label{eqn-h}
\eb
which is solved either by $h = 0$ (isotropic case), or by a time evolving $h$
which is decreasing by a rate set by the average expansion parameter $H \,$.

We want to study the possible background solutions for this system under the assumption of initial 
slow roll of $\phi \,$. Let us first study the approximate case in which $V$ is constant, and the inflaton is
at rest (vacuum energy). The first and third of~(\ref{bg.system}) are then solved by
\begin{equation}
H = \frac{\sqrt{V}}{\sqrt{3} \, M_p} \, \frac{{\rm e}^{\,2 \sqrt{3 V}\,t / M_p} + 1}{{\rm e}^{\,2 \sqrt{3 V}\,t / M_p} - 1}
\;\;\;,\;\;\; h = \pm \, \frac{2 \, \sqrt{V}}{M_p} \, \frac{{\rm e}^{\sqrt{3 V} \, t / M_p}}{{\rm e}^{\,2 \sqrt{3 V}\,t / M_p} - 1}
\;\;,\;\; V \; {\rm constant}
\label{solvconst}
\end{equation}
We note the presence of two distinct branches of solutions, characterized by 
either positive or negative values of $h \,$. In the following, we refer to them as 
the {\it positive} and the {\it negative branch}, respectively.

We see that the metric has a singularity at $t=0 \,$. Close to the singularity, the system approaches a Kasner~\cite{kasner} ``vacuum solution'', where the presence of the source ($V$) can be neglected. The line element of  
Kasner geometries is
\begin{equation}
d s_{\rm Kasner}^2 = - d t^2 + t^{2 \alpha} d x^2 + t^{2 \beta} d y^2 + t^{2 \gamma} d z^2
\label{kasner}
\end{equation}
where the exponents satisfy the two properties  
 $\alpha+\beta+\gamma=\alpha^2+\beta^2+\gamma^2=1\,$. In the present case ($\beta = \gamma$), we can invert the relations~(\ref{defHh}), to find, close to the singularity,
\begin{eqnarray}
H_a \rightarrow \frac{1}{t} \;\;,\;\; H_b \rightarrow 0 
&,&\;\; {\rm positive \; branch} \nonumber\\
H_a \rightarrow - \frac{1}{3 \, t} \;\;,\;\; H_b \rightarrow \frac{2}{3 \, t} \;\;&,&\;\; {\rm negative \; branch}
\end{eqnarray}
Integrating these rates reproduces the line element (\ref{kasner}) with $\alpha = 1 ,\, \beta=\gamma=0$ in the positive branch, and $\alpha=- 1/3 ,\, \beta=\gamma=2/3$ in the negative branch. 

We note that the singularity is only a coordinate one for the positive branch, while it is a real one for the negative branch. This can be seen by computing the curvature invariants in both branches; while $R$ is finite at $t =0$ in both cases, the product $R_{\mu \nu \rho \sigma} R^{\mu \nu \rho \sigma}$ is finite in the positive branch, while it diverges in the negative one. It is possible to perform a change of coordinates in the $t,x$ plane for which the metric is regular in the region corresponding to $t=0 \,$. This would select a new time, which could be used for the quantization of the perturbations. However, we prefer to choose the time $t$ for the quantization, since the background is constant on fixed $t$ surfaces.
As we will see, this procedure allows us to choose an adiabatic initial vacuum for all modes.

We also see from (\ref{solvconst}) that, for both branches, the system
quickly reaches isotropy on a timescale $M_p / \sqrt{V} \,$ (which
parametrically coincides with the inverse of the Hubble rate due to
the vacuum energy), after which one quickly approaches the de Sitter values $H =
\sqrt{V/3 M_p^2} \,,\, h = 0 \,$. We note that, in the negative branch, the
scale factor of the anisotropic direction experiences a bounce, since
it contracts close to the singularity, while at late times it expands
with the asymptotic de Sitter rate.

Let us now return to the complete system~(\ref{bg.system}). The only
restriction that we impose is that the inflaton $\phi$ is initially in
a slow roll regime. Since $h$ decreases very rapidly, we need to start from
a strong anisotropy in order to have a measurable deviation from the
standard (isotropic) inflationary results. To do so, we choose 
$\vert h \vert \gg \sqrt{V} / M_p$ at the start. In this regime, the average rate $H$ is much
greater than in the isotropic case, so that, due to the higher Hubble
friction, the inflaton is initially rolling even more slowly than in
the standard case. For this reason, the results obtained above for
constant $V$ describe very accurately also the evolution of the
complete system during the isotropization stage.

\begin{figure}[ht]
\centering
\includegraphics[width=9cm,clip,angle=-90]{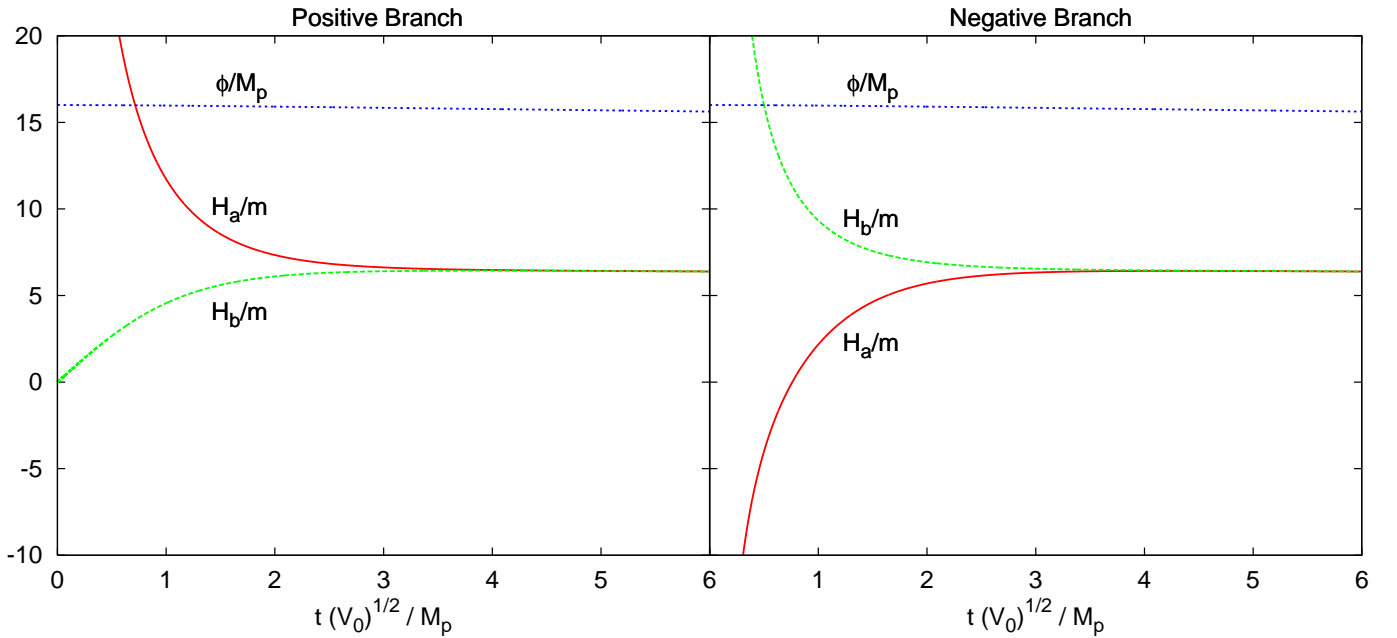}
\caption{\label{fig.bck} Time evolution of the two expansion rates
$H_a$ (red-solid) and $H_b$ (green-dashed), in units
of $m$, and of the inflaton  (blue-dotted), in units of $M_p$, for a
choice of chaotic inflaton potential $V=m^2\phi^2/2$. The left panel refers to the
positive branch, while the right one to the negative branch. $V_0$ denotes the
initial potential of the inflaton, with $\phi_0 = 16\, M_p$ in this
example. Notice that for the negative branch $H_a$ starts from negative values, indicating that this direction
is initially contracting.
}
\end{figure}

The degree of accuracy can be checked for any given inflaton potential, by solving the system~(\ref{bg.system}) with an expansion series in time. For massive chaotic inflation, with $V = m^2 \phi^2 / 2 \,$, we find
\begin{eqnarray}
&& H = \frac{1}{3 \, t} \left[ 1 + \frac{m^2 \, \phi_0^2 \, t^2}{2 \, M_p^2} + {\rm O} \left( m^4 \, t^4 \right) \right] \;\;,\;\; h = \pm \frac{1}{\sqrt{3} \, t} \left[ 1 - \frac{m^2 \, \phi_0^2 \, t^2}{4 \, M_p^2} + {\rm O} \left( m^4 \, t^4 \right) \right] \;\;,\;\; \nonumber\\
&& \phi = \phi_0 \left[ 1 - \frac{m^2 \, t^2}{4} + {\rm O} \left( m^4 \, t^4 \right) \right]
\label{early}
\end{eqnarray}
(these expansions do not assume $m \ll M_p$, although this is the regime of phenomenological interest). In figure \ref{fig.bck} we plot the time evolution of the two expansion rates $H_a$ and $H_b$, and of the
inflaton field; the left panel refers to the
positive branch, while the right one to the negative branch; $V_0$ denotes the
initial potential of the inflaton. The evolutions shown (obtained by a numerical integration of the system
(\ref{bg.system})), are in perfect
agreement with the above discussion. The background has an initial strong anisotropy (we choose $h_0 = \pm 10^6 \, m$ in the two branches, respectively), which is rapidly damped away, and it is then followed by a stage of standard isotropic inflation. Since the inflaton is nearly static in the early anisotropic stage, its initial value controls the duration of the isotropic inflationary expansion. The value $\phi_0 = 16 \, M_p \,$ chosen in the figure leads to about $60$ e-folds of standard inflation.


\section{Cosmological Perturbations}
\label{pert}

We now compute the perturbations about the backgrounds presented in
the previous Section. 

It is useful to recall first the standard results. In presence of a
single scalar field (the inflaton) in a homogeneous and isotropic FRW
universe one finds three physical perturbations. These three modes
are always decoupled from each other, due to the symmetry properties
of the background. The actual computation can be found, among other works, in the
comprehensive review~\cite{mfb}.  The two polarizations of the tensor
mode are denoted by $h_\times$ and $h_+$. The scalar degree of freedom
is instead encoded in the so called Mukhanov--Sasaki variable $v$
\cite{musa}, which is in turn related to the comoving curvature
perturbation ${\cal R}$ as
\begin{equation}
{\cal R} = \frac{H}{a \, \dot{\phi}} \, v
\end{equation}
These quantities are gauge invariant (they do not change under
coordinate transformations), and ${\cal R}$ represents the
gravitational potential on comoving hypersurfaces where the inflaton
field is homogeneous (this property defines these hypersurfaces).

In the remainder of this Section we discuss the anisotropic case.
We do so in three parts. In the first
part, we show how the isotropic computation can be extended to the
anisotropic background presented in the previous Section. We also
outline the procedure to further generalize this computation to
arbitrary anisotropic spaces. In the second part we discuss the initial
conditions for the perturbations both in the positive and negative
branch. We conclude with the computation and the discussion of the
curvature power spectrum.

\subsection{Anisotropic background}
\label{pert1}

The number of the physical modes can be obtained by a simple counting. We start 
from $10$ perturbations in the (symmetric) metric, plus $1$ perturbation of the
inflaton field. Of these $11$ perturbations, $4$ are gauge modes,
which can be set to zero once we completely fix the freedom of general
coordinate reparameterization. Of the remaining modes, $4$ are
nondynamical (namely, they enter with at most one time derivative in the
action), and $3$ physical. A
convenient method to extract the $3$ physical modes is to choose a set
of conditions for the perturbations which fixes the gauge completely
(thus removing the $4$ gauge modes) without eliminating any $\delta
g_{0 \mu}$ mode. These modes are nondynamical, as can be most easily
seen in the ADM formalism~\cite{adm}, where the metric elements
$g_{00}$ and $g_{0i}$ are written as Lagrange multipliers. We can
therefore integrate them out of the action, as we explicitly 
show in Appendix \ref{appB3} and \ref{appB4}.~\footnote{This is analogous to what happens in
electromagnetism. There, one starts from the $4$ components of the
vector potential. One degree of freedom is eliminated by fixing the $U
\left( 1 \right)$ gauge, while the $A_0$ component is non dynamical
(it can be fixed by Gauss' law). One is then left with the $2$
physical polarizations of the massless photon. Clearly, one can also
choose different gauges; for example, a gauge often chosen in the isotropic
gravity computation is the longitudinal one, which removes the $\delta g_{0i}$ modes. In
general, however, the nondynamical modes will be linear combinations
of the remaining perturbations, and it may be less straightforward to
find them.} We remark that this counting does not assume any symmetry of the 
$4$ dimensional background. Therefore, the procedure that we have just outlined can
be applied to more general geometries than the simple Bianchi $I$ considered here.

We first performed the computation in conformal time $\eta$, defined as
\begin{equation}
d s^2 = a^2 \left( \eta \right) \left( - d \eta^2 + d x^2 \right) + b^2 \left( \eta \right) \left( d y^2 + d z^2 \right)
\label{back}
\end{equation}
We explain the reason for this choice in the next subsection, where we also discuss how
different choices of time can affect the  quantization procedure.

We denote the physical perturbations about this background by $H_\times ,\, H_+ $,
and $V$, with the understanding that, when the background becomes
isotropic, each quantity becomes the corresponding lower case
perturbation introduced in the previous subsection. The explicit definition 
of the physical modes in terms of the starting
metric and inflaton perturbations is presented in
appendix~\ref{appB}. Here we only outline the equations satisfied by
these modes, while in the next subsection we give their initial
conditions. The form of these equations is the same for both
branches. The mode $H_\times$ is always decoupled from the other two,
as a consequence of the residual isotropy in the $y-z$ plane. The
other two modes are coupled to each other. Denoting by prime a
derivative with respect to conformal time, we find
\begin{eqnarray}
&& H_\times'' + \omega_\times^2 \, H_\times = 0 \nonumber\\
&& \left( \begin{array}{c} V \\ H_+ \end{array} \right)'' + 
\brn{ \begin{array}{cc}
\omega_{11}^2 & \omega_{12}^2 \\
\omega_{12}^2 & \omega_{22}^2
\end{array} }  \, 
\left( \begin{array}{c} V \\ H_+ \end{array} \right) = 0\,.
\label{evol}
\end{eqnarray}
The explicit expressions for the frequency elements are given in eqs.~(\ref{vec2d}) and (\ref{matome}). 

Correspondingly, the quadratic actions for these modes are formally identical to the action of oscillators
in Minkowski space with time evolving frequencies. This generalizes the Mukhanov-Sasaki~\cite{musa} computation, valid in the FRW case: we can then proceed to the quantization of the modes, and provide their initial conditions (see the next subsection).

As the background becomes isotropic, $b \rightarrow a$, we find 
\begin{eqnarray}
&& \omega_\times^2 ,\, \omega_{22}^2 \rightarrow k^2 - \frac{a''}{a} \nonumber\\
&& \omega_{11}^2 \rightarrow k^2 - \frac{z''}{z} \;\;\;,\;\;\; z \equiv \frac{a^2 \phi'}{a'} \nonumber\\
&& \omega_{12}^2 \rightarrow 0
\label{omegiso}
\end{eqnarray}
where $k$ is the comoving momentum. These are the standard evolution equations for the FRW case. All the modes decouple, and the equations for the two tensor polarizations become identical.

\subsection{Initial conditions} 
\label{subinitial}

The goal of this subsection is to set the initial conditions for the perturbations at asymptotically early initial times.
 The key quantities are the frequencies (\ref{vec2d}) and (\ref{matome}), which we evaluate at asymptotically early times through the background relations~(\ref{early}).

Again, it is useful to start by summarizing the standard isotropic computation. The physical modes $h_\times ,\, h_+ ,\, v$ are those that diagonalize the quadratic action for the perturbations. In conformal time, defined as
\begin{equation}
d s^2 = a^2 \left( \eta \right) \left[ - d t^2 + d x^2 + d y^2 + d z^2 \right]
\label{stdconf}
\end{equation}
the action for each of these modes (after Fourier transforming the spatial coordinates) is of the form
\begin{eqnarray}
S_{(2)} &=& \frac{1}{2} \int d \eta \, d^3 k \left[ \vert \delta' \vert^2 - \omega^2
\vert \delta \vert^2 \right] \nonumber\\
\omega^2 &=& k^2 - a^2 {\cal F} \left( \eta \right)
\label{azd}
\end{eqnarray}
where $\delta$ denotes any of the modes, and ${\cal F}$ is a time evolving
function which is of the order of $H^2 $. The term $a^2 {\cal F}$ is
exponentially small at asymptotically early times. The net result is
that at early times the action of the mode (once written in conformal
time) approaches the one of a simple harmonic oscillator in Minkowski
spacetime, with the frequency equal to the comoving momentum $k \,$.
In the Minkowski case we would have
\begin{equation}
\delta_{\rm Mink.} = \frac{{\rm e}^{-i k \eta}}{\sqrt{2 k}}
\end{equation}
for a choice of positive frequency modes. In the cosmological
context, this is replaced by the adiabatic vacuum
\begin{equation}
\delta_{\rm in} \simeq \frac{{\rm e}^{-i \int^\eta d \eta' \, \omega}}{\sqrt{2 \omega}}
\label{adia}
\end{equation}
which is a solution of the evolution equation as long as the frequency
is adiabatically changing, $\omega' \ll \omega^2 \,$. The fact that
$k$ dominates $\omega$ at early times, can be also restated as $k / a
\gg H \,$. Going backwards in time, the universe is nearly
exponentially contracting during inflation, and the physical momentum
$p = k / a$ becomes the dominant quantity for the evolution of the
mode.

Let us now discuss the anisotropic situation. We separate the comoving momentum ${\bf k} = {\bf k}_1 + {\bf k}_2 \,$, where ${\bf k}_1$ denotes the component along the privileged direction $x $, and ${\bf k}_2$ the component in the $y-z$ plane. We also denote by $k_1$ and $k_2$ the magnitudes of the two components. The magnitude of the physical momentum is therefore
\begin{equation}
p^2 = \frac{k_1^2}{a^2} + \frac{k_2^2}{b^2}
\end{equation}

For the positive branch, during the anisotropic phase, $a \propto \, t \,$,
while $b$ is nearly constant. As we go backwards in time, only the $x$
direction is ``squeezing'', while the other two become
frozen. Correspondingly, $k_1 / a \gg k_2 / b$ at sufficiently early
times, and we therefore expect that only the component of the momentum
in that direction sets the initial condition. To be able to set an
initial vacuum as in the isotropic case, we therefore use the scale
factor $a$ in the definition of the conformal time, cf
eqs.~(\ref{back}) and~(\ref{stdconf}). Indeed, by inserting the early
time asymptotic behaviours (\ref{early}) into the various frequency
elements of eq.~(\ref{evol}), we find
\begin{equation}
\omega_\times^2 = k_1^2 + {\rm O} \left( m^2 t^2 \right) \;\;\;,\;\;\;
\brn{ \begin{array}{cc}
\omega_{11}^2 & \omega_{12}^2 \\
\omega_{12}^2 & \omega_{22}^2
\end{array} }  =
\brn{ \begin{array}{cc}
k_1^2 + {\rm O} \left( m^2 t^2 \right) & {\rm O} \left( m^2 t^2 \right) \\
{\rm O} \left( m^2 t^2 \right)   & k_1^2 + {\rm O} \left( m^2 t^2 \right)
\end{array} }  
\label{freqearly}
\end{equation}
so that $k_1$ always becomes dominant provided we go sufficiently
early in time.\footnote{The only exception is when $k_1$ is strictly
zero, that is when the momentum of the mode lies in the $y-z$ plane.
This is a set of measure zero in the integral (\ref{corrcllmm}) for
the correlators $C_{\ell \ell' m m'}$, so we can disregard this
exceptional case. Notice also that we used physical time to indicate
the residual entries in the frequency element, while the evolution
equations~(\ref{evol}) are in conformal time.} Moreover, we see that
the modes $H_+$ and $V$ decouple also at asymptotically early times,
so that we can set the initial conditions directly on them (rather
than on the modes which would diagonalize the frequency
matrix). Therefore, as long as
\begin{equation}
\frac{\omega_\times'}{\omega_\times^2} \;,\;\; \frac{\omega_{11}'}{\omega_{11}^2} \;,\;\; \frac{\omega_{22}'}{\omega_{22}^2} \;,\;\; \frac{\omega_{12}^2}{\omega_{11}^2} \;,\;\; \frac{\omega_{12}^2}{\omega_{22}^2} \,\ll\, 1
\label{adia2}
\end{equation}
we can consistently start from the adiabatic vacuum~(\ref{adia}) for the perturbations. This always happens, provided we go sufficiently early in time (the smaller $k_1$ is, the earlier we have to go in time to satisfy these conditions).

In appendix \ref{appC1}, we study whether other choices of time are possible to set the initial conditions. In the isotropic case, the conformal time is related to the physical one by $d t = a \, d \eta$. The most obvious generalization is to consider some arbitrary power of the two scale factors in the definition of the conformal time. We therefore discuss conformal times $\tau$ related to $\eta$ by
\begin{equation}
d \eta = a^{-2 \alpha} \, b^{-2 \beta} d \tau
\label{etatau}
\end{equation}
where $\alpha$ and $\beta$ are two constant parameters. We show in appendix \ref{appC1} that this leads to an adiabatic evolving frequency only if $\alpha = 0$. Since $b$ is constant at asymptotically early times, we conclude that the time $\eta$ used here is the only possible one for the quantization, up to a trivial constant rescaling. In particular, this rules out the most ``symmetric'' choice $d t = a^{1/3} \, b^{2/3} \, d \tau$, where the scale factors along the three coordinates are used with an equal weight.

Let us now turn to the negative branch. In conformal time $\eta$, the frequency elements appearing in eq.~(\ref{evol}) are the same as for the positive branch (namely, eqs.~(\ref{vec2d}) and (\ref{matome})).
However, the early time asymptotics are now different (due to the fact that we now take the negative solution for $h$ in the asymptotic eq.~(\ref{early})):
\begin{equation}
\omega_\times^2 = 
a^2 \left[ - \frac{5}{9 t^2} + \frac{k_2^2}{b^2} + {\rm O} \left( 1 \right) \right]
\;\;\;,\;\;\;
\brn{ \begin{array}{cc}
\omega_{11}^2 & \omega_{12}^2 \\
\omega_{12}^2 & \omega_{22}^2
\end{array} }  = a^2
\left[ \brn{ \begin{array}{cc}
\frac{4}{9 t^2} + \frac{k_2^2}{b^2} & 0 \\
0   & \frac{4}{9 t^2} + \frac{k_2^2}{b^2}
\end{array} }  + {\rm O} \left( 1 \right) \right]
\label{freqearly2}
\end{equation}
where we recall that in the negative branch $a \propto t^{-1/3} ,\; b \propto t^{2/3}$ at asymptotically early times.

The situation is now significantly worse than for the positive branch. Firstly, the frequencies are not adiabatically evolving at early times:
\begin{equation}
\frac{\vert \omega_\times \vert'}{\omega_\times^2} \rightarrow \frac{4}{\sqrt{5}} \;\;\;\;\;\;,\;\;\;\;\;\;
\frac{\omega_{11}'}{\omega_{11}^2} \;,\; \frac{\omega_{22}'}{\omega_{22}^2} \rightarrow - 2
\end{equation} 
For this reason, we are not able to start from an adiabatic vacuum at early times, as we could do for the positive branch. Secondly, the $2$d vector is tachyonic.

As we show in appendix \ref{appC2}, it is not possible to find a conformal time in which the frequency of the $2$d scalars is initially adiabatically evolving. The situation is somewhat better for the $2$d vector modes; by performing a time redefinition of the form~(\ref{etatau}), with $\alpha = 2 \left( 1 + \beta \right)$, the frequency $\omega_\times^2$ approaches a constant negative value in the asymptotic past. In this case, we may choose to start from an adiabatic vacuum (using the absolute value of $\omega_\times^2$). However, the fact that $\omega_\times^2$ is negative makes the mode diverge. To see this, we note that, for $\alpha = 2 \left( 1 + \beta \right)$, the transformation between the physical time $t$ and the comoving time $\tau$ reads
\begin{equation}
d t = a \, d \eta = a^{-2 \alpha + 1} \, b^{- 2 \beta} \, d \tau = c \, t \, d \tau
\end{equation}
where $c$ is a positive constant. Integrating this relation, we see that the initial singularity ($t=0$ in physical time) occurs at $\tau = - \infty$. Since $\omega_\times^2$ approaches a negative constant as $\tau \rightarrow - \infty$, the $2$d vector experiences a tachyonic growth which makes it diverge. More appropriately, we conclude that the $2$d vector becomes non-linear, and the perturbative computation breaks down. Modes with different scales and directions of momenta are coupled at the nonlinear level; this will likely ``isotropize'' the effective Hubble drag felt by the non-linear modes. Intuitively this would lead to a feedback resulting in the suppression of the their growth. Nonetheless, the breakdown of the linearized computation (together with the lack of adiabaticity in the $2$d scalar sector) prevents us from making trustable and firm predictions for the temperature anisotropies in the negative branch,
and for this reason we do not consider this branch further in the
remainder of the analysis.

\subsection{Curvature power spectrum}
\label{subcurva}

\begin{figure}[th]
\centering
\includegraphics[width=12cm,clip]{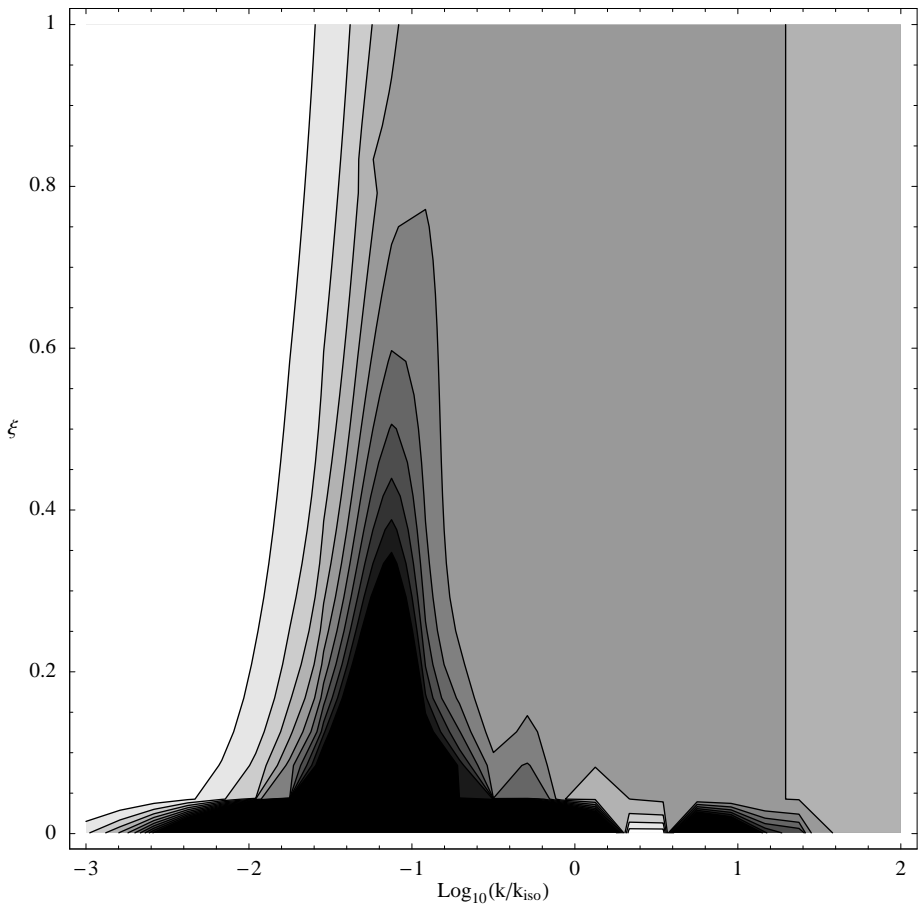}
\caption{Power spectrum of the comoving curvature perturbation ${\cal R}$ in the inflationary model $V = m^2 \,\phi^2 /2 $. The inflaton field starts with $\phi = 16 \,M_p$, and is evolved until the moment shown in the figure, when $\phi = M_p$. Modes with $k>k_\iso$ leave horizon during the later isotropic stage of inflation. The standard result is recovered for these modes.}
\label{fig.scal}
\end{figure}

\begin{figure}[ht]
\centering
\includegraphics[width=12cm,clip]{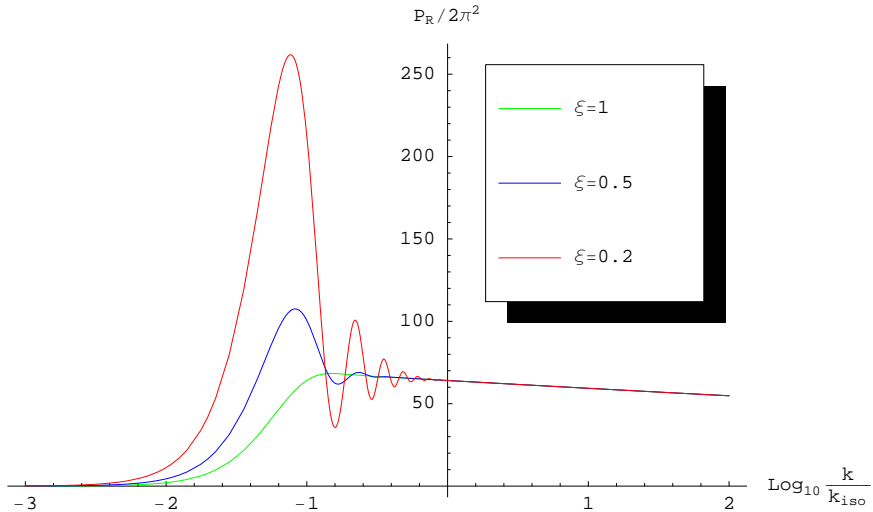}
\caption{Sections of the power spectrum (same evolution as in the previous figure) at fix values of $\xi \,$. We see that the power spectrum becomes isotropic (no $\xi$ dependence) at large momenta; at lower momenta it presents some oscillations, whose amplitude increases as $\xi$ decreases (this leads to the $1/\xi$ divergency mentioned in the main text). Finally, the power spectrum is suppressed as $k \rightarrow 0 \,$.
}
\label{fig.fixxi}
\end{figure}

We want to compute the power spectrum for the (late time) scalar mode. According to the discussion of the previous subsection, we restrict our attention to the positive branch. We proceed as follows. We fix an initial value for the inflaton ($\phi_0 = 16 \, M_p$ in all the cases studied in this work, so to have about $60$ e-foldings of isotropic inflation), and we scan over several values for the two components $k_{1,2}$ of the comoving momentum of the mode. We choose a sufficiently early time such that the conditions~(\ref{adia2}) are satisfied, and we set the initial conditions for the modes and their derivatives as given by the adiabatic solution~(\ref{adia}). We then numerically evolve the evolution equations~(\ref{evol}) for the coupled $\left\{ V ,\, H_+ \right\}$ system, until a final time for which the background is isotropic 
(so that $V \equiv v ,\, H_+ \equiv h_+$) and the modes of interest are well outside the horizon. In figure \ref{fig.scal} we show a contour plot with the value of $k^3 \vert v \vert^2$ at the end of inflation. This quantity is the curvature power spectrum,  up to an overall normalization factor. The horizontal axis in the figure is the magnitude $k = \sqrt{k_1^2 + k_2^2}$ of the comoving momentum, in units of a reference momentum $k_{\rm iso}$. The latter is defined as the comoving momentum of the modes which exit the horizon when the universe becomes isotropic; namely, $k_{\rm iso} \equiv a H$ at $t = t_{\rm iso} \,$, which we conventionally define to be the time at which $h = 10^{-3} m$. The vertical axis is instead the cosine of the angle between the comoving momentum and the $x-$axis (namely, $k_1 = k \, \xi \,$). In figure \ref{fig.fixxi} we show instead some sections of the power spectrum at fixed values of $\xi \,$.

Modes with $k \gg k_{\rm iso}$ leave the horizon during the isotropic
phase; at early times, their frequency changes adiabatically, and, as
a consequence, they evolve according to the adiabatic
solution~(\ref{adia}). As the background becomes isotropic, the
frequency $\omega$ becomes the standard one, so that 
the mode adiabatically evolves to the 
adiabatic solution that one would have 
also found if the
background had always been isotropic. For this reason, we expect to
recover an isotropic power spectrum (no $\xi$ dependence) in this
limit. This behaviour is manifest in the results shown. Larger scale
modes ($k \la k_{\rm iso}$) are instead more sensitive to the background
evolution in the anisotropic phase, and we expect nonstandard
results in this regime. Indeed, at any fixed $\xi \,$, the power
spectrum exhibits an oscillatory behaviour for $k \la k_{\rm iso}$, and it
then sharply decreases for $k \rightarrow 0 \,$.

We also observe that the power spectrum increases at low $\xi \,$. The numerical results indicate that the 
the power spectrum actually diverges as $1/\xi$ there:
\begin{equation}
P \left( k ,\, \xi \right) \simeq \frac{{\cal P} \left( k \right)}{\xi} \;\;\;\,\;\;\;
{\rm for} \; 0 < \xi< {\bar \xi} \left( k \right)
\label{growp}
\end{equation}
The region where this happens shrinks to smaller and smaller size as $k$ grows (namely, 
${\bar \xi} \left( k \right)$ is a decreasing function of $k$ in the region we have probed),
in agreement with the fact that the power spectrum approaches the isotropic result 
for $k \rightarrow \infty \,$. Nonetheless, the growth of the power spectrum 
takes place at all values of $k$ we have computed, provided $\xi$ is sufficiently small.

Although we do not have an analytical solution of the
system~(\ref{evol}), the large power at small $\xi$ is most probably due to the initial
normalization $\vert v_{\rm in} \vert \propto k_1^{-1/2} \propto
\xi^{-1/2} \,$. In the isotropic case, modes with very small values of
$k$ start with a large amplitude $\vert v_{\rm in} \vert \propto
k^{-1/2} \,$, but this does not lead to a large power spectrum $P
\propto k^3 \vert v \vert^2$, since the prefactor $k^3$ is small in
this regime. For the anisotropic case, the problem arises when $k$ is
finite, but $\xi \rightarrow 0 \,$. In this regime, $k_1 \ll k_2 \,$,
so that the physical momentum is almost lying in the $y-z$ plane in
the later isotropic regime. However, the initial condition for the
mode is set at sufficiently early times, for which $k_1 / a_{\rm in}
\gg k_2 / b_{\rm in} \,$. At the initial time, the mode has the large
amplitude $\vert v_{\rm in} \vert \propto \xi^{-1/2} \,$, while the
prefactor $k^3$ in the late time power spectrum $k^3 \vert v \vert^2$ equals 
$k_2^3 \,$, which is not
suppressed in the $\xi \rightarrow 0 \,$ limit.

The divergency in the power spectrum results in a logarithmic
divergency in the $C_{\ell \ell' m m'}$ correlators, due to the fact
that the associated Legendre polynomials approach a finite constant
value for $\xi \rightarrow 0$ in the angular integral of
eq.~(\ref{corrcllmm}), while the power spectrum diverges there. We do
not expect that this singularity will actually occur, but it rather
indicates the breakdown of the linearized computation for the
metric/inflaton perturbations. This means that the precise value of
the correlators is sensitive to the nonlinear dynamics of the modes,
which cannot be accounted for in the present linearized computation.

In addition to this, we note that the growth~(\ref{growp}) occurs at
small $k_1 \,$. We recall that the early time frequency of the modes
is $\omega^2 = k_1 + {\rm O } \left( t^2 \right) \,$, see
eq.~(\ref{early}), and that the metric is singular at $t=0
\,$. The adiabaticity condition, $\omega' / \omega^2 \ll 1$ is satisfied
by all modes at sufficiently early times. However, the smaller $k_1$
is, the closer one needs to go to the singularity for the adiabaticity
condition to hold. Suppose that the background solution can be trusted
only from some time on. Then, we can consistently start from the
adiabatic vacuum only for modes with sufficiently high momentum
$k_1$. The value of the modes with lower momentum may be simply
provided as an (arbitrary) initial condition, or may be controlled by
the additional dynamics (possibly, by additional degrees of freedom),
which may be relevant at earlier times.

To conclude, we expect that the singularity in the power spectrum is
actually absent once the nonlinear effects and/or the complete early
dynamics of the system are taken into account. In either case, however, this
means that additional inputs are needed to provide firm predictions
for the temperature anisotropies in this simple model.


\section{Discussion}
\label{conc}
The main result of this work is the extension of the computation of
primordial perturbations in a FRW geometry to more general
backgrounds.  We have done this in detail, only for the simplest case
of a Bianchi $I$ model with residual isotropy in two spatial directions,
and with one scalar inflaton field: as in the isotropic case, there
are three physical modes. However, in contrast to this case, two of
the modes are coupled to each other already at the linearized level,
due to the fact that the background has less symmetries. At late
times, when the background becomes isotropic, one of these two modes
becomes the scalar perturbation, while the other one becomes one of the
two perturbations of the tensor mode. The key
step in the computation is to completely fix the gauge for the
perturbations without removing the $\delta g_{0 \mu}$ entries. Such modes 
are nondynamical, and can be readily
integrated out of the quadratic action of the perturbations. 
In this way, one is left with only the physical degrees of freedom.
It is straightforward to extend this
computation to a general Bianchi $I$ model, without the residual $2$d
spatial isotropy. In this case, all the three modes will be coupled to
each other at the linearized level. Moreover, this procedure can be
performed also for more general backgrounds, or for cases in which
more fields are present.

Such anisotropic backgrounds have a potentially very interesting
phenomenology; for instance the early time coupling between the scalar
and tensor modes produces an off diagonal correlation between the CMB
modes (eq. \ref{corrcllmm}), and, possibly a nonstandard
tensor-to-scalar ratio. Moreover the two polarizations of the tensor
modes behave differently at early times, and acquire different values. 
Therefore, a future detection of the tensor perturbations
(for instance, through the B polarization modes) can provide extremely
useful information to test this possibility.

The computation that we have outlined is a necessary first step for
such studies, since it must be done in order to find the physical
modes, and to fix their initial conditions. In standard isotropic
inflation, the frequency of the modes is controlled by their physical momentum
$p$ at early times, which is varying adiabatically due to the
expansion of the universe, $\omega' / \omega^2 \ll 1 \,$. It is
customary to start from the adiabatic vacuum (the immediate
generalization of the de Sitter Bunch--Davies \cite{buda} vacuum), which is a
good solution of the equations of motion as long as the adiabaticity
condition holds.~\footnote{The choice of the initial vacuum is however
object of intense debate, see for instance~\cite{mabr}, mostly due to
the fact that the physical momentum is trans-Planckian close to the
initial singularity, and unknown UV physics may leave some imprint in
the mode.} Due to this fact, the final result (a nearly scale
invariant spectrum of the perturbations, in perfect agreement with
observations) is insensitive to the ``initial time'' at which the
initial vacuum is set, provided that the mode is deeply inside the
horizon ($p \gg H$) at this time.

Models which go beyond the standard FRW geometry have additional
parameters. To have some predictive power, one should hope that 
their phenomenology depends on as few inputs as
possible. For instance, if we find that 
also in this case the frequency 
of the physical perturbations varies
adiabatically at early times, then the problem of setting the initial
vacuum is no worse than for standard inflation. We have seen
that this is not the case for the simple Bianchi $I$ backgrounds that
we have studied here. The main reason is that the anisotropy parameter
$h$ (the difference of the two expansion rates) decreases very rapidly
at the onset of inflation. Clearly, this geometry is continuously connected
to FRW, and one can take $h$ arbitrary small at the initial time. In
this case, however, one recovers standard FRW inflation plus negligible
corrections. If one hopes instead to find sizeable deviations from
the isotropic case, the initial value of $h$ must be large. In
this work we have studied the early evolution of the model starting from
$\vert h \vert \gg \sqrt{V} / M_p$ (where $V$ is the potential energy of the inflaton),
assuming that the inflaton is in a slow roll regime also during the
anisotropic phase. This leads to two distinct possibilities: in one
case, two dimensions become static at asymptotically early times; in
the other case, the anisotropic direction is initially
contracting. In the second case, the frequency of two of the modes is
not adiabatically evolving at early times, while for the other one is actually
tachyonic. In the first case, we do find an initial adiabatic regime; however, the
amplitude of the modes increases and eventually diverges as their
momentum is more and more aligned towards the non expanding
directions.

As we have mentioned, this divergence is merely an indication of the
breakdown of the linear approximation taken in our treatment. It is
probable that nonlinear dynamics will stop the growth of the modes at
this stage. However, even if this is the case, the predictive power of the
model is reduced greatly, since the much simpler 
linearized treatment is invalid.
There may be other physical reasons
why the run-off growth in the mode may not occur.  A better behaving 
vacuum may result if all the directions are expanding at asymptotically 
early times. For a Bianchi $I$ geometry, even dropping the assumption
of residual $2$d isotropy, this cannot be achieved
if the inflaton is initially in a slow roll regime (this can be understood from the
$00$ Einstein equation; close to the initial singularity the three expansion
rates cannot be all positive, if the energy density of the inflaton remains finite).
Alternatively, it is possible that the Bianchi $I$ geometry is only the final stage of the
anisotropic phase, and that additional degrees of freedom are relevant
at earlier times. Both these possibilities require additional input with
respect to the minimal set-up studied here.

One may also study completely different anisotropic
models from the start. For instance, it may be very interesting to
study models which evade Wald's theorem on the isotropization of
Bianchi geometries~\cite{wald}. This can 
happen if one modifies Einstein gravity, for instance 
through the Kalb--Ramond action of
string theory \cite{nemanja}, or through quadratic curvature
invariants \cite{barrow}. In such case, one can find attractor
solutions characterized by anisotropic inflationary expansion. This
may avoid the run-off growth problem found in the model we have studied,
since the difference $h$ between the expansion rates decreases slower, or not at all,
in these models, and therefore does not need to be extremely large
in the asymptotic past. The challenge for such models is to  
find solutions for which the anisotropy decays at late times, or,
alternatively, can be kept at a very small but controllable
level. This condition itself might lead to smaller, subdominant effects
in the data which are yet to be observed. 

We have taken a top-down approach in suggesting a model-oriented
origin for the broken isotropy observed in the CMB. We have examined a
particular form of anisotropy in the inflating universe but the
treatment presented here can be readily generalized to other forms of
anisotropy. If the intriguing CMB anomalies are verified to be of
cosmological origin through more accurate, future observations,
including the polarization data, then they may well indicate a
departure from the simplest picture of inflation. In that case models
such as the one presented here offer a minimal modification of the
inflationary paradigm which could predict the anomalous
correlations required to explain the the data.

\vspace{1cm} {\bf Note added: } While this work was being completed,
we became aware of the work \cite{uzan}, which also develops a
formalism for the computation of anisotropies in Bianchi I
models. Although the approach is rather different, and a different conformal time
is used in \cite{uzan}, we have verified 
that the evolution for the canonical modes obtained in \cite{uzan}
agree with ours in the limit in which the
background has a $2$d residual isotropy. The work \cite{uzan} does not
study the background solutions for the model at the extent 
done here, and it therefore does
not discuss the initial conditions for the perturbations, nor the
resulting power spectrum and CMB temperature anisotropies. We also
note that some of the results for the perturbations in the anisotropic
model discussed here had also been summarized in \cite{ours}.

\vspace{1cm}

{\bf \large Acknowledgements}

\bigskip

\noindent We thank John D. Barrow, Nemanja Kaloper, Lev Kofman, and Keith A. Olive for very useful discussions. We also thank an anonymous referee for relevant comments on the first version of this manuscript. The work of M.P. was partially supported by  the DOE grant DE-FG02-94ER-40823.


\appendix
\def\theequation{\thesection.\arabic{equation}}
\setcounter{equation}{0}

\begin{center}
\section*{Appendixes}
\end{center}

\bigskip

\appendix
\def\theequation{\thesection.\arabic{equation}}
\setcounter{equation}{0}

\section{Explicit computation of $C_{\ell \ell' m m'}$}
\label{appA}

We want to prove the result (\ref{corrcllmm}) appearing in the main
text. A full treatment of the line of sight calculation of CMB
anisotropies \cite{zaldarriaga} can be found in
\cite{dodelson}. Inverting the relation (\ref{dt}), the
expression~(\ref{cllmm}) becomes
\begin{equation}
  C_{\ell\ell' m m'} = \vev{a^{\,}_{\ell m}a^\star_{\ell'm'}} = \int
  d\Omega_\vp d\Omega_\vpp \vev{\delta T(\vp,\eta_0,{\bf x}_0)\delta T(\vpp,\eta_0,{\bf x}_0)} Y_{\ell m}^\star(\vp)Y^{\,}_{\ell' m'}(\vpp),
\end{equation}
The temperature anisotropy is related to the primordial curvature perturbation as in~(\ref{dt2}). The statistical correlation is therefore encoded in the primordial power spectrum~(\ref{power}) of the curvature perturbation. This gives
\begin{equation}
 C_{\ell\ell' m m'} = \int \frac{d^3{\bf k}}{(2\pi)^3} \frac{1}{k^3}
 P({\bf k})\int
  d\Omega_\vp d\Omega_\vpp \Delta(k,\vk\cdot\vp,\eta_0)\Delta^\star(k,\vk\cdot\vpp,\eta_0)Y_{\ell m}^\star(\vp)Y^{\,}_{\ell' m'}(\vpp),
\end{equation}
Note that the {\it directional} dependence of the radiation transfer functions
$\Delta(k,\vk\cdot\vp,\eta_0)$ at conformal time today $\eta_0$ is
restricted to the cosine of the angle subtended by the plane wave unit
vector ${\bf \hat k}$ and the unit vector in the direction on the sky
${\bf \hat p}$. 
Since the universe is already isotropic when the initial conditions for the
Einstein-Boltzmann system are set,  the standard
line of sight calculation can be used in the calculation of the
multipole expanded transfer functions $\Delta_\ell(k,\eta_0)$, which
depend only on the magnitude of the wave vector. 

We expand the transfer functions as in~(\ref{deltadec}), and we
decompose the Legendre polynomials entering in (\ref{deltadec}) into
spherical harmonics
\begin{equation}
  P_L(\vk\cdot\vp) = \frac{4\pi}{(2L+1)} \sum_{M=-L}^L Y^{\star}_{L
  M}(\vk)Y_{L M}(\vp).
\end{equation}
so that the correlator can be rewritten as
\begin{eqnarray}
   C_{\ell\ell' m m'} &=& \int \frac{d^3{\bf k}}{(2\pi)^3}
 \frac{(4\pi)^2}{k^3} P({\bf k})
 \sum_{LL'MM'}(-i)^{L-L'}\Delta_L(k,\eta_0)\Delta^{\star}_{L'}(k,\eta_0)
 Y^{\star}_{L M}(\vk)Y_{L' M'}(\vk)\nonumber\\ &&\times \int
 d\Omega_\vp d\Omega_\vpp Y_{\ell m}^\star(\vp)Y^{\,}_{\ell'
 m'}(\vpp)Y_{LM}^{\,}(\vp)Y^{\star}_{L' M'}(\vpp).
\end{eqnarray}
It is then straightforward to compute the angular integrals, given the
orthonormality of the spherical harmonics. This gives
\begin{equation}
   C_{\ell\ell' m m'} = (-i)^{\ell-\ell'}\int \frac{d^3{\bf k}}{(2\pi)^3}
 \frac{(4\pi)^2}{k^3} P({\bf k})
 \Delta_\ell(k,\eta_0)\Delta^\star_{\ell'}(k,\eta_0)
 Y^\star_{\ell m}(\vk)Y_{\ell' m'}(\vk).
\end{equation}
Note that this expression is valid for {\it any} form of anisotropic
initial curvature power spectrum if the universe is isotropized by the
time of reheating.

By assumption, the power spectrum in the model considered here is
axially symmetric around the anisotropic $z-$axis. We therefore use
polar coordinates in momentum space, where $\theta$ is the angle
between ${\bf k}$ and the $z-$axis. Due to the residual symmetry, the
integral over the second angular coordinate $\phi$ is trivial. More
specifically, we expand
\begin{equation}
  Y_{\ell m} (\theta,\phi) = e^{im\phi}
  \sqrt{\frac{(2\ell+1)(\ell-m)!}{4\pi(\ell+m)!}} P_\ell^m( \xi ),
\end{equation}
where $\xi = \cos \theta \,$, and we use the fact that
\begin{equation}
  \int_0^{2\pi} d\phi Y^{\,}_{\ell m} (\theta,\phi)Y^\star_{\ell' m'}
  (\theta,\phi) = 2\pi \delta_{mm'}
  \frac{1}{4\pi}\sqrt{\frac{(2\ell+1)(2\ell'+1)(\ell-m)!(\ell'-m)!}{(\ell+m)!(\ell'+m)!}}P_\ell^m(\cos\theta) P_{\ell'}^m(\cos\theta),
\end{equation}
to obtain eq.~(\ref{corrcllmm}) of the main text. If the power
spectrum is isotropic, the first angular integral is also carried out
trivially and using the property
\begin{equation}
\int_{-1}^1 d \xi \, P_\ell^m \left( \xi \right) P_{\ell'}^m \left( \xi \right) = \frac{2 \left( \ell + m \right)!}{\left( 2 \ell + 1 \right) \left( \ell - m \right)!} \delta_{\ell \ell'},
\end{equation}
we recover the standard expression~(\ref{cllmmiso}).

\section{Explicit computation of the perturbations }
\label{appB}

In this appendix we perform the explicit computation of the
perturbations about the background geometry~(\ref{back}). 
Since this is a linearized computation, we ignore the nonlinear
coupling between different modes. We can therefore fix a comoving
momentum ${\bf k}$, and study the evolutions of the modes having that
momentum. The computation is exhaustive as long as we can solve the
problem for any arbitrary value of ${\bf k}$. Due to the residual
symmetry in the $y-z$ plane, we can fix $k_z = 0$ without any loss of
generality. More appropriately, we denote by $k_1$ the component of
the momentum along the anisotropic $x$ direction, and by $k_2$ the
component in the orthogonal plane. We then choose the $y-z$
coordinates such that the orthogonal momentum lies on the $y$
direction. In coordinate space, this amounts to considering modes
which depend on $x$ and $y$ only.~\footnote{In our computation we
disregard the cases where the $x$ and $y$ dependence are
trivial. Namely, we always assume that both $k_1$ and $k_2$ are
different from zero. These two cases are of zero measure in the
integral (\ref{corrcllmm}) and can therefore be disregarded.} The most
general perturbations of the metric can be then written as
\begin{equation}
g_{\mu \nu} =\left(\begin{array}{llll}
- a^2 \left(1+2 \, \Phi\right) & a \, \partial_1 \chi&a \, \partial_2 B & b^2 \, B_3\\
& a^2 \left(1-2\,\Psi\right) & b^2 \, \partial_1 \partial_2 {\tilde B} & b^2 \,\partial_1\tilde B_3\\
& & b^2 \left(1-2\, \Sigma + 2 \, \partial_2^2 E \right) & b^2 \, \partial_2 E_3\\
& & & b^2 \left(1-2\,\Sigma \right)
                 \end{array} \right)\,.
\label{metric}                 
\end{equation}
In addition, there is the perturbation of the inflaton field $\varphi = \phi + \delta \phi \,$. 

In the isotropic computation, the modes are classified into tensor, vector, or scalar modes, according to how they transform under $3$d spatial rotations. This is particularly useful, since, due to the isotropy of the background, perturbations belonging to different groups are decoupled at the linearized level \cite{bardeen, mfb}. A $3$d scalar has one degree of freedom. A $3$d vector has one spatial index; however, due to the transversality condition ($\partial_i V_i = 0$), it has two degrees of freedom. A $3$d tensor has two spatial indices, and it is required to be symmetric, transverse ($\partial_i h_{ij} = 0$), and traceless. Therefore, it also has $2$ degrees of freedom. Altogether, the $10 + 1$ initial perturbations (of the metric and of the inflaton field), are classified into $5$ scalar, $2$ vector, and $1$ tensor modes.~\footnote{We stress that this is only the first step in the computation, which is dictated by mathematical convenience. Not all of these modes are physical; altogether, $4$ degrees of freedom are gauge modes, and $4$ other ones are nondynamical. See the discussion in Subsection~\ref{pert1}.}

Here we do the same, but we classify the perturbations according to how they transform under $2$d rotations (since modes will be decoupled only if they transform differently under $2$d rotations in the $y-z$ plane, due to the restricted isotropy of the background). A $2$d scalar has one degree of freedom. A $2$d vector has $2$ spatial indices, and is required to be transverse; therefore, it has one degree of freedom. Contrary to the previous case, $2$d tensors instead do not exist, since there are no degrees of freedom left, once we impose the mode to be symmetric, transverse, and traceless. Therefore, the $10 + 1$ initial perturbations are classified into $8$ scalar and $3$ vector modes. 

We conclude this discussion with a note on our notation. A $2d$ vector in the $y-z$ plane is in general indicated with $V_i \,$, where $i=2 ,\, 3$. The transversality condition is then $\partial_2 V_2 + \partial_3 V_3 = 0 \,$. However, as mentioned before eq.~(\ref{metric}), we can impose that the modes do not depend on the third coordinate, without any loss of generality. Therefore, we can set $V_2 = 0 \,$, and the degree of freedom of the $2$d vector is explicitly encoded into its component $V_3 \,$. This justifies the choice of the $2$d vector modes $B_3 ,\, {\tilde B}_3 ,\, E_3$ appearing in the metric~(\ref{metric}). The parameterization of the $2$d scalars entering in~(\ref{metric}) has been chosen for computational convenience.

\subsection{Gauge choice}
\label{appB1}

We can then proceed by either constructing gauge invariant perturbations, or by choosing a gauge which completely removes the gauge freedom (in the standard case, this is typically done by the choice of the longitudinal gauge). The two procedure are completely equivalent, due to gauge invariance. We choose the second option, by setting
\begin{equation}
\delta g_{12} = \delta g_{23} = \delta g_{22} = \delta g_{33} = 0
\label{ourgauge}
\end{equation}
Unlike the more conventional gauge choices, we chose not to remove any perturbation entering in the $g_{0 \mu}$ metric elements. As we mentioned in Subsection \ref{pert1}, these modes are nondynamical~\cite{adm}. Therefore, with this gauge choice, we know from the start that the nondynamical modes of~(\ref{metric}) are $\Phi ,\, \chi ,\, B ,\,$ and $B_3 \,$. This modes will be then integrated out from the action of the perturbations (see below).

It is straightforward, although tedious, to show that this choice (i) can be made, and (ii) completely fixes the gauge freedom. This is true both for the isotropic ($b = a$) and anisotropic background. Under the general coordinate transformation $x^\mu \rightarrow x^\mu + \xi^\mu$, the metric transforms as
\begin{equation}
\delta g_{\mu \nu} \rightarrow \delta g_{\mu \nu}-\,^{(0)}g_{\mu\nu,\alpha}\xi^\alpha -\,^{(0)}g_{\mu\alpha}\xi^\alpha_{\;\;,\nu} -\,^{(0)}g_{\alpha\nu}\xi^\alpha_{\;\;,\mu}
\label{eqtransf}
\end{equation}

We parametrize $\xi^\mu = \left(\xi^0,\partial_1 \xi^1,\partial_2 \xi, \xi_3\right)$, so that $\xi_3$ is
the $2$d vector degree of freedom, while the other three degrees of freedom are $2$d scalars.
Eq.~(\ref{eqtransf}) then explicitly gives
\begin{eqnarray}
\Phi&\rightarrow&\Phi- \frac{a'}{a} \xi^0 - \xi^{0 '}\,,
\nonumber\\
\chi&\rightarrow&\chi+ a \left( \xi^0 - \xi^{1 '} \right)\,,
\nonumber\\
B&\rightarrow& B+ a \, \xi^0 - \frac{b^2}{a} \xi'\,,
\nonumber\\
\Psi&\rightarrow& \Psi+ \frac{a'}{a} \, \xi^0+\partial_1^2 \xi^1\,,
\nonumber\\
\tilde B&\rightarrow& \tilde B-\frac{a^2}{b^2}\xi^1- \xi\,,
\nonumber\\
\Sigma &\rightarrow& \Sigma +\frac{b'}{b} \, \xi^0\,,
\nonumber\\
E &\rightarrow& E - \xi\,,
\nonumber\\
B_3 &\rightarrow& B_3 -b \,\xi_3'\,,
\nonumber\\
\tilde B_3 &\rightarrow& \tilde B_3 - \xi_3\,,
\nonumber\\
E_3 &\rightarrow& E_3-\xi_3\,.
\nonumber\\
\delta \phi &\rightarrow& \delta \phi - \phi' \, \xi^0
\label{transf}
\end{eqnarray}
(for completeness, in the last line we have indicated how the perturbation of the inflaton transforms). 
Our gauge choice corresponds to ${\tilde B} = \Sigma = E = E_3 = 0 $. It is easy to see that
this is possible for one and only one choice of the parameters $\xi^\mu \,$.

\subsection{Comparison with the standard modes in the isotropic limit ($b = a$)}
\label{appB2}

It is instructive to see how the conventional modes (tensor and scalar
modes of~\cite{mfb}) rewrite in this gauge, once the background
becomes isotropic. This will help us interpreting our results. To do
so, we start from the conventional mode parameterization, without $3$d
vectors (which are not supported during inflation) and with the scalar
modes written in the longitudinal gauge. In the momentum space (where
the comparison is easier) we have:
\begin{equation}
g_{\mu \nu} = a^2 \left( \begin{array}{llll}
- 1 - 2 {\bar \Phi} & 0 & 0 & 0 \\
& 1 - 2 {\bar \Psi} + \frac{k_2^2}{k^2} {\bar h}_+ & - \frac{k_1 k_2}{k^2} {\bar h}_+ & \frac{k_2}{k} {\bar h}_\times \\
& & 1 - 2 {\bar \Psi} + \frac{k_1^2}{k^2} {\bar h}_+ & - \frac{k_1}{k} {\bar h}_\times \\
& & & 1 - 2 {\bar \Psi} - {\bar h}_+
\end{array} \right)
\label{long}
\end{equation}
where $k^2 = k_1^2 + k_2^2 \,$.

In terms of the general decomposition~(\ref{metric}) (written in the isotropic limit $b=a$) the expression~(\ref{long}) rewrites
\begin{eqnarray}
&& \Phi = {\bar \Phi} \;\;,\;\; \chi = B = B_3 = 0 \nonumber\\
&& \Psi = {\bar \Psi} - \frac{k_2^2}{2 \, k^2} \, {\bar h}_+ \;\;,\;\;
{\tilde B} = \frac{{\bar h}_+}{k^2} \;\;,\;\;
E = - \frac{2 \, k_1^2 + k_2^2}{2 \, k^2 \, k_2^2} \, {\bar h}_+ \;\;,\;\;
\Sigma = {\bar \Psi} + \frac{{\bar h}_+}{2} \nonumber\\
&& {\tilde  B}_3 = \frac{i \, k_2}{k_1 \, k} \, {\bar h}_\times \;\;,\;\;
E_3 = - \frac{i \, k_1}{k \, k_2} \, {\bar h}_\times
\label{gentolong}
\end{eqnarray}

It is straightforward to write down how the final (and properly
normalized) gauge invariant perturbations $v ,\, h_+ ,\,$ and
$h_\times$ appear in the longitudinal gauge~\cite{mfb}:
\begin{equation}
v = a \left[ \delta {\bar \phi} + \frac{a \phi'}{a'} {\bar \Psi} \right] \;\;,\;\;
h_+ = \frac{M_p}{\sqrt{2}} \, a \, {\bar h}_+ \;\;,\;\;
h_\times = \frac{M_p}{\sqrt{2}} \, a \, {\bar h}_\times
\label{gaugeinv}
\end{equation}
where $\delta {\bar \phi}$ is the perturbation of the inflaton in the longitudinal gauge.

We then transform from the longitudinal gauge to our gauge~(\ref{ourgauge}). Namely, we start from the
perturbations~(\ref{metric}) in the longitudinal gauge~(\ref{gentolong}), and perform an infinitesimal change of coordinates $x^\mu \rightarrow x^\mu + \xi^\mu$, such that the transformed modes satisfy the conditions that specify our gauge, ${\tilde B} = \Sigma = E = E_3 = 0$ (the explicit transformations of the modes are given by eqs.~(\ref{transf}) in the isotropic limit $b = a ,\, b' = a'$). This can be achieved for a unique choice of the infinitesimal transformation parameters $\xi^\mu \,$:
\begin{equation}
\xi^0 = - \frac{a}{a'} \left( {\bar \Psi} + \frac{{\bar h}_+}{2} \right) \;\;,\;\;
\xi^1 = \frac{2 \, k_1^2 + 3 \, k_2^2}{2 \, k^2 \, k_2^2} \, {\bar h}_+ \;\;,\;\;
\xi = - \frac{2 \, k_1^2 + k_2^2}{2 \, k^2 \, k_2^2} \, {\bar h}_+ \;\;,\;\;
\xi_3 = - \frac{i \, k_1}{k \, k_2} \, {\bar h}_\times
\label{parinf}
\end{equation}

Starting from the longitudinal gauge~(\ref{gentolong}), and performing the infinitesimal transformations~(\ref{transf}) with the parameters~(\ref{parinf}), the dynamical modes $\delta \phi ,\, \Psi ,\, {\tilde B}_3$
in our gauge become
\begin{equation}
\delta \phi = \delta {\bar \phi} + \frac{\phi' \, a}{a'} \left( {\bar \Psi} + \frac{{\bar h}_+}{2} \right) \;\;,\;\;
\Psi = - \frac{k^2}{k_2^2} \, {\bar h}_+ \;\;,\;\;
{\tilde B}_3 = \frac{i \, k}{k_1 \, k_2} \, {\bar h}_\times
\end{equation}
We invert these relations to express the modes in the longitudinal gauge in terms of our modes. 
We then insert the resulting expressions into eqs.~(\ref{gaugeinv}); in this way, we find how our modes combine into the gauge invariant combinations $v ,\, h_+ ,\, h_\times $:
\begin{eqnarray}
a \left[ \delta \phi + \frac{k_2^2}{2 k^2} \frac{a \, \phi'}{a'} \, \Psi \right] = v \;\;,\;\;
- a \, \frac{M_p}{\sqrt 2} \, \frac{k_2^2}{ k^2} \Psi = h_+ \;\;,\;\;
- i a \, \frac{M_p}{\sqrt{2}} \, \frac{k_1 \, k_2}{k} \, {\tilde B}_3 = h_\times
\label{inter}
\end{eqnarray}

\subsection{$2$d vectors}
\label{appB3}

We now compute the quadratic action for the $2$d vectors. We insert the decomposition~(\ref{metric}) into the action~(\ref{action}), ignoring the $2$d scalars. We also set $E_3 = 0$, according to the gauge choice~(\ref{ourgauge}), and we expand the action up to second order in the remaining $2$d modes $B_3$ and ${\tilde B}_3 \,$. This gives
\begin{eqnarray}
S_{(2)}^{\rm 2d vec} = \frac{M_p^2}{4} \int d\eta \, d^3 x ~b^2\, \left[ \frac{b^2}{a^2} \left(
\tilde{B}'^2_{3 ,1}  + B^2_{3,1} - 2 \, B_{3 ,1}\,\tilde{B}'_{3,1}   \right) + 
 B_{3 ,2}^2 - \tilde{B}_{3,1,2}^2 \right]
 +\mathcal{B}^{\rm vec}_1\,,
\label{vecpar1}
\end{eqnarray}
where the boundary terms are
\begin{eqnarray}
\mathcal{B}^{\rm vec}_1 &=& M_p^2\,\int d\eta\,d^3 x \left[ \mathcal{F}_\eta\,' + \partial_1 \mathcal{F}_x +\partial_2 \mathcal{F}_y \right]\,,\nonumber\\
\mathcal{F}_\eta &=& \frac{b^4}{2\,a} \left[ -\left(H_a +2\, H_b\right) B_3^2  -\frac{1}{b^2}\left(\frac{b^2}{a} \tilde{B}_{3,1}^2\right)'\right] \,,\nonumber\\
\mathcal{F}_x &=& \frac{1}{a} \left(\frac{b^4}{a}\,B_3\,\tilde{B}_{3,1}\right)'-\frac{b^4}{2\,a^2}\left(B_3^2\right)_{,1}\,,\nonumber\\
\mathcal{F}_y &=& \frac{b^2}{2} \left( \tilde{B}^2_{3 ,1} -  B^2_{3}\right)_{,2}
\end{eqnarray}
($H_a = \dot{a}/a $ and $ H_b = \dot{b}/b$ are the Hubble rates with respect to physical time, see eqs.~(\ref{defhahb})). Next, we Fourier transform along the spatial coordinates. We can see from the action~(\ref{vecpar1}) that the mode $B_3$ is nondynamical, and it can be integrated out (as we discussed 
in~\ref{pert1}, the modes $\delta g_{0 \mu}$ are nondynamical). Extremizing the Fourier transformed action with respect to $B_3^\star$ gives
\begin{equation}
B_3 = \frac{p_1^2}{p^2} \, \tilde{B}'_{3} \,,
\end{equation}
where we have used for shortness the components of the physical momentum $p_1 = k_1 / a$ and $p_2 = k_2 / b$, and the total magnitude $p = \sqrt{p_1^2 + p_2^2}$.  We insert this value for $B_3$ back in the Fourier transformed action. The resulting action depends only on $\tilde{B}_3$ and its first derivative. We can cast it in the form
\begin{equation}
S_{(2)}^{\rm 2d vec}  = \frac{1}{2} \, \int d \eta \, d^3 k \, \left[ \vert H_\times' \vert^2 - \omega_\times^2 \, \vert H_\times \vert^2 \right] + \mathcal{B}^{\rm vec}_1 + \mathcal{B}^{\rm vec}_2
\label{az2dv}
\end{equation}
where
\begin{eqnarray}
H_\times &\equiv& \frac{M_p}{\sqrt{2}} \, \frac{b \, k_1 \, k_2}{\sqrt{k_1^2 + \frac{a^2}{b^2} k_2^2}} \, {\tilde B}_3  \nonumber\\
\omega_\times^2 &\equiv& k_1^2 + a^2 \, \left[ \frac{k_2^2}{b^2} - H_a^2 - H_b^2 + \frac{\dot{\phi}^2}{2 M_p^2} + \left( H_a - H_b \right)^2 \frac{k_1^2 \left( k_1^2 + 4 \frac{a^2}{b^2} k_2^2 \right)}{\left( k_1^2 + \frac{a^2}{b^2} k_2^2 \right)^2} \right] 
\label{vec2d}
\end{eqnarray}
and where the newly introduced boundary term is
\begin{equation}
\mathcal{B}^{\rm vec}_2 = \frac{1}{2}\int d\eta \, d^3 k \left\{  \left[ - H_b + \left( H_a - H_b \right) \frac{p_2^2}{p^2} \right] \, a \, \vert H_\times \vert^2\right\}'
\end{equation}

Namely, the action~(\ref{az2dv}), disregarding the boundary terms, has been written in terms of the 
canonically normalized combination $H_\times$. This mode coincides with the standard mode $h_\times$ in the isotropic limit, as can be seen by comparing its definition with the third of (\ref{inter}) when $b = a \,$ (the modes actually differ by an unphysical phase). Its equation of motion, appearing in eq.~(\ref{evol}) of the main text, is immediately obtained from (\ref{az2dv}).

\subsection{$2$d scalars}
\label{appB4}

We now compute the quadratic action for the $2$d scalars. The computation proceeds analogously to the one of previous subsection, although it is more involved due to the presence of a larger number of modes. We start from the metric~(\ref{metric}), in the gauge (\ref{ourgauge}), and we now ignore the $2$d vectors. We also include the perturbation of the inflaton field in the present computation. We insert this in the action~(\ref{action}) and we expand it to quadratic order in the perturbations. We find
\begin{eqnarray}
S_{(2)}^{\rm 2d sca} &=&\frac{M_p^2}{4} \int d\eta\, d^3x \, a^2b^2\,\left\{ 
\frac{2}{M_p^2}\left(\frac{1}{a^2} \delta\phi'^2-\frac{1}{a^2} \delta\phi_{,1}^2-\frac{1}{b^2} \delta\phi_{,2}^2\right)\right. \nonumber\\
&&+\frac{4}{b^2}\left[ -\frac{\dot{\phi}}{M_p^2} \delta\phi + \left(H_a+H_b\right) \Phi +\left(H_a-H_b\right) \Psi +\frac{1}{a} \Psi'\right]_{,2}B_{,2}\nonumber\\
&&+\frac{1}{a^2b^2} \left(B_{,1,2}-\chi_{,1,2}\right)^2-\frac{4\,\dot{\phi}}{M_p^2 a}\left(\Psi+\Phi\right) \delta\phi' + \frac{4\,V'}{M_p^2} \left(\Psi-\Phi\right) \delta \phi\nonumber\\
&&+\frac{4}{a^2} \left(2\,H_b \Phi -\frac{\dot{\phi}}{M_p^2}\delta\phi\right)_{,1}\,\chi_{,1}
-\frac{4}{b^2} \Phi_{,2}\Psi_{,2}-\frac{8}{a}\,H_b \Phi\Psi'\nonumber\\
&&\left. - \frac{2\,V''}{M_p^2} \, \delta\phi^2 +2\, \left[ \frac{\dot{\phi}^2}{M_p^2}-2\, H_b \left(2\,H_a +H_b\right) \right] \Phi^2\right\}
+\mathcal{B}^{\rm sca}_1
\end{eqnarray}
where we stress that $\dot{\phi}$ is the derivative of the background inflaton with respect to physical time, and where the boundary terms are
\begin{eqnarray}
\mathcal{B}^{\rm sca}_1 &=& \frac{M_p^2}{2}\,\int d\eta\,d^3 x \left[ \mathcal{F}_\eta\,' + \partial_1 \mathcal{F}_x +\partial_2 \mathcal{F}_y \right]\,,\nonumber\\
\mathcal{F}_\eta &=&  a\,b^2\left[\left(H_a +2\, H_b\right) \left(3\,\Phi^2 +2\,\Phi \Psi- \frac{1}{a^2}\chi_{,1}^2 -\frac{1}{b^2} B_{,2}^2\right) \right.\nonumber\\
&&\left. -H_a \Psi^2 +\frac{2}{a} \left(\Phi - \Psi\right) \Psi'-\frac{2}{a^2} \chi_{,1} \Phi_{,1} -\frac{2}{b^2} B_{,2} \Phi_{,2} \right]\,,\nonumber\\
\mathcal{F}_x &=& b^2 \, \left[2\,\left(2\,H_b \chi_{,1} +\Phi_{,1} +\frac{1}{a} \chi_{,1}'\right)\left(\Phi-\Psi\right) -\frac{2}{b^2}  \, B_{,2} \, B_{,1,2} +\frac{2}{a} \chi_{,1} \Phi'\right] \,,\nonumber\\
\mathcal{F}_y &=& 2\,a^2 \left\{ \left[ \left(H_a+H_b\right)B_{,2} +\Phi_{,2} +\Psi_{,2} +\frac{1}{a}B_{,2}'\right]\left(\Phi+\Psi\right)\right.\nonumber\\
&&\left. +\frac{1}{a^2}\left[\left(B-\chi\right)_{,1,2} \chi_{,1} + \left( \chi_{,1,1}+a\left(\Phi+\Psi\right)' \right)B_{,2}\right] \right\}
\end{eqnarray}
We now find that the modes $\Phi ,\, \chi $, and $B$ are nondynamical (they are the $\delta g_{0 \mu}$ modes), while the two remaining modes, $\Psi$ and $\delta \phi$ are dynamical. We can then integrate out the nondynamical perturbations analogously to what we did for the $2$d vectors, and obtain an action in terms of the two dynamical modes only.

After Fourier transforming the spatial component, extremizing the action wrt $\chi^\star$ gives
\begin{equation}
\chi = B + \frac{2\,b^2}{k_2^2} \left(\frac{\dot{\phi}}{M_p^2}\,\delta\phi-2\,H_b \Phi\right) \,,
\end{equation}
Inserting this back into the Fourier transformed action gives
\begin{eqnarray}
S_{(2)}^{\rm 2d sca} &=& M_p^2 \int d\eta \, d^3k\,a^2\,b^2 \left\{ \frac{1}{2\, a^2 \, M_p^2} \vert \delta\phi'\vert ^2 - \frac{1}{2 M_p^2} \left[p_1^2+p_2^2+V''+\frac{2\,p_1^2\,\dot{\phi}^2}{p_2^2 M_p^2}\right]\vert\delta\phi\vert^2 \right. \nonumber\\
&& +\left[ \frac{\dot{\phi}^2}{2\,M_p^2} -H_b \left(2\,H_a +H_b\left(1+\frac{4\,p_1^2}{p_2^2}\right)\right)\right] \vert\Phi\vert^2 \nonumber\\
&& \!\!\!\!\!\!\!+\frac{1}{2}\left(\left[\left(2\,H_b p_1^2 +\left(H_a +H_b\right)p_2^2\right)\Phi - \frac{\dot{\phi}}{M_p^2}\left(p_1^2+p_2^2\right)\,\delta\phi +p_2^2\frac{b}{a^2} \left(\frac{a}{b} \Psi\right)'\right]B^\star \right.\nonumber\\
&&\;\;\;+\left[-p_2^2 \Psi +\frac{4\,H_b p_1^2 \dot{\phi}}{p_2^2 M_p^2} \delta\phi - \frac{\dot{\phi}}{M_p^2 a} \,\delta\phi'-\frac{2\,H_b}{a} \Psi'-\frac{V'}{M_p^2} \delta\phi\right]\Phi^\star\nonumber\\
&& \left.\left.  \;\;\;-\frac{1}{M_p^2} \left( \frac{\dot{\phi}}{a} \,\delta\phi' - V'\,\delta\phi\right)\Psi ^\star + {\rm h.c.} \right)\right\}+\mathcal{B}^{\rm sca}_1 \,.
\label{scapar1}
\end{eqnarray}
We extremize this action wrt $B^\star$. The resulting equation of motion can be written as an equation for $\Phi$
\begin{equation}
\Phi = \frac{p_2^2}{H_a \, p_2^2 + H_b \left( 2 p_1^2 + p_2^2 \right)}\left[ \frac{\dot{\phi}}{M_p^2} \left(1+\frac{p_1^2}{p_2^2}\right)\delta\phi - \frac{b}{a^2}\left(\frac{a}{b} \Psi\right)'\right]\,.
\end{equation}
When we insert this back into~(\ref{scapar1}), also $B$ disappears from the action. The resulting action can be finally cast in the form
\begin{equation}
S = \frac{1}{2} \, \int d \eta \, d^3 k \, \left[ 
\vert V ' \vert^2 + \vert H_+' \vert^2 - \left( V^* ,\, H_+^* \right) \, \Omega^2 \left( \begin{array}{c} V \\ H_+ \end{array} \right) \right] +\mathcal{B}^{\rm sca}_1 +\mathcal{B}^{\rm sca}_2 
\label{az2ds}
\end{equation}
where the canonically normalized combinations are
\begin{equation}
V \equiv b \left[ \delta \phi + \frac{p_2^2 \, \dot{\phi}}{H_a \, p_2^2 + H_b \left( 2 p_1^2 + p_2^2 \right)}
\Psi \right] \;\;,\;\;
H_+ \equiv \frac{\sqrt{2} \, b \, M_p \, p_2^2 \, H_b}{H_a \, p_2^2 + H_b \left( 2 p_1^2 + p_2^2 \right)} \Psi 
\label{sca2d}
\end{equation}
We see by comparing with the first two of (\ref{inter}) that these modes reduce to the analogous (lower case) ones in the isotropic limit (up to an unphysical phase for $H_+$ \,). 

The term $\Omega^2$ entering in the action is the square frequency matrix
\begin{eqnarray}
\label{matome}
\Omega^2 &=& \left( \begin{array}{cc}
\omega_{11}^2 & \omega_{12}^2 \\
\omega_{12}^2 & \omega_{22}^2
\end{array} \right) \\ \nonumber\\ 
\nonumber\\
\left(\frac{\omega_{11}}{a}\right)^2 &=& \left(p_1^2 +p_2^2 -2\,H_a \, H_b 
+ \frac{3\,\dot \phi^2}{2\,M_p^2} + \frac{2\,H_a}{H_b}\frac{\dot\phi^2}{M_p^2} - 
\frac{1}{2\,H_b^2}\frac{\dot\phi^4}{M_p^4} + \frac{2}{H_b} \frac{\dot\phi\,V'}{M_p^2}+ 
V''\right)
\nonumber\\
&& +\frac{p_2^2\left(H_a-H_b\right)}{\left[2\,H_b\,p_1^2+\left(H_a+H_b\right)p_2^2\right]}
\frac{\dot\phi}{M_p}\left[-\frac{4\,\dot\phi}{M_p} - \frac{2\,H_a}{H_b}\frac{\dot\phi}{M_p}
+\frac{1}{H_b^2}\frac{\dot\phi^3}{M_p^3}-\frac{2}{H_b}\frac{V'}{M_p}\right.
\nonumber\\
&&\quad\quad\quad\quad\quad\quad\quad\quad\quad\quad
\left.-\frac{p_2^2\left(H_a-H_b\right)}{\left[2\,H_b\,p_1^2+\left(H_a+H_b\right)p_2^2\right]}
\frac{\dot\phi}{M_p}\left(1+\frac{1}{2\,H_b^2}\frac{\dot\phi^2}{M_p^2}\right)\right]
\nonumber\\
\left(\frac{\omega_{22}}{a}\right)^2  &=& p_1^2+p_2^2-2\,H_a\,H_b+\frac{\dot\phi^2}{2\,M_p^2}
\nonumber\\
&&\quad\quad\quad\quad
+\frac{p_2^2\left(H_a-H_b\right)^2}{\left[2\,H_b\,p_1^2+\left(H_a+H_b\right)p_2^2\right]}
\left[4\,H_b - \frac{p_2^2\left(2\,H_b^2+\frac{\dot\phi^2}{M_p^2}\right)}
{\left[2\,H_b\,p_1^2+\left(H_a+H_b\right)p_2^2\right]}\right]
\nonumber\\
\left(\frac{\omega_{12}}{a}\right)^2&=& 
\frac{\sqrt{2} \, p_2^2\left(H_a-H_b\right)}{\left[2\,H_b\,p_1^2 + \left(H_a+H_b\right)p_2^2\right]}
\left[-3\,H_b\frac{\dot\phi}{M_p} + \frac{1}{2\,H_b}\frac{\dot\phi^3}{M_p^3} - \frac{V'}{M_p}
\right.\nonumber\\
&&\left.\quad\quad\quad\quad\quad\quad\quad\quad\quad
-\frac{p_2^2\left(H_a-H_b\right)}{\left[2\,H_b\,p_1^2 + \left(H_a+H_b\right)p_2^2\right]}
\frac{\dot\phi}{M_p}\left(H_b+\frac{1}{2\,H_b}\frac{\dot\phi^2}{M_p^2}\right)\right]
\nonumber
\end{eqnarray}

The equation of motion for the $2$d scalars, appearing in eq.~(\ref{evol}) of the main text, is immediately obtained from (\ref{az2ds}).

The second boundary term is
\begin{eqnarray}
\mathcal{B}^{\rm sca}_2 &=& \int d\eta\,d^3k\,  \left[ \left( V^* ,\, H_+^* \right) \,
\left(\begin{array}{ll}
 \mathcal{F}_{VV} & \mathcal{F}_{VH} \\
 \mathcal{F}_{VH} &\mathcal{F}_{HH}
\end{array}
 \right) \left( \begin{array}{c} V \\ H_+ \end{array} \right)\right]'\,,\nonumber\\
\mathcal{F}_{VV} &=& -\frac{a}{2}\left(H_b +\frac{\left(p_1^2+p_2^2\right)\dot{\phi}^2}{\left(H_a \, p_2^2 + H_b \left( 2 p_1^2 + p_2^2 \right)\right)M_p^2}\right)\,,\nonumber\\
\mathcal{F}_{VH} &=& \frac{a}{2\sqrt{2}\,M_p} \left[\frac{V'}{H_b} -2\,\dot{\phi}\frac{\left( 2\,H_b \,p_1^4 +H_a p_2^2 \left(p_1^2 -p_2^2\right)\right)}{p_2^2 \left(H_a \, p_2^2 + H_b \left( 2 p_1^2 + p_2^2 \right)\right)}\right] \,,\nonumber\\
\mathcal{F}_{HH} &=& a\left\{ -\frac{\dot{\phi}\,V'}{4\,H_b^2\,M_p^2} +\frac{4\,H_b\,p_1^2 +H_a\,p_2^2}{4\,M_p^2\,H_b^2\,p_2^2} \, \dot{\phi}^2 +\frac{2\,H_b\,p_1^2 +\left(H_a+H_b\right)p_2^2}{4\,H_b^2}\right.\nonumber\\
&& +\frac{1}{p_2^2\left[H_a \, p_2^2 + H_b \left( 2 p_1^2 + p_2^2 \right)\right]} \left[ -2\,H_b \left(H_a-H_b\right)p_1^4 \right.\nonumber\\
&&\left.\left. - \left(2\,H_a^2 +H_a\,H_b -2\,H_b^2\right) p_1^2\,p_2^2-\frac{H_a}{2\,H_b} \left(H_a^2 +2\, H_a \,H_b-H_b^2\right)p_2^4\right]\right\}\nonumber\\
\end{eqnarray}

\section{Dependence of the frequency, and of the adiabaticity condition on the choice of conformal time}
\label{appC}

In Appendix \ref{appB}, we compute the frequencies of the perturbations in conformal time $\eta \,$, defined in eq.~(\ref{back}). Here, we study how the action of the modes changes when we use another conformal time $\tau \,$. 

Rather than studying the $2$d vector and scalar sector separately, we can give a more compact presentation by considering $N$ canonically normalized fields $X_i$ (after Fourier transforming the spatial coordinates), with the action
\begin{equation}
S = \frac{1}{2} \int d \eta \, d^3 k \left[ \frac{d X^\dagger}{d \eta} \, \frac{d X}{d \eta} - X^+ \, \Omega_\eta^2 \, X \right] \;\;\;\;\;\;,\;\;\; X \equiv \left( \begin{array}{c} X_1 \\ X_2 \\ \dots \\ X_N \end{array} \right)
\label{azx}
\end{equation}
where $\Omega_\eta^2$ is a hermitian matrix (in the present application, $N=1$ and $2$ for the $2$d vector and the $2$d scalar sector, respectively). 

Under the time redefinition
\begin{equation}
d \eta \equiv f^{-2} \left( \tau \right) \, d \tau
\label{newtime}
\end{equation}
this action becomes
\begin{equation}
S = \frac{1}{2} \int d \tau d^3 k \left[ f^2 \, \frac{d X^\dagger}{d \tau} \, \frac{d X}{d \tau} - f^{-2} \, X^+ \, \Omega_\eta^2 \, X \right]
\end{equation}
With this new time variable, the fields $X_i$ are no longer canonical. We can however introduce new canonical fields
\begin{equation}
Y_i \equiv f \, X_i
\label{newfield}
\end{equation}
In terms of these fields, the action reads
\begin{equation}
S = \frac{1}{2} \int d \tau \, d^3 k \left[ \frac{d Y^\dagger}{d \tau} \, \frac{d Y}{d \tau} - Y^+ \, \Omega_\tau^2 \, Y \right] + {\cal B} 
\label{azy}
\end{equation}
where the new square frequency matrix is
\begin{equation}
\Omega_\tau^2 \equiv \frac{1}{f^4} \, \Omega_\eta^2 - \frac{1}{f} \, \frac{d^2 f}{d \tau^2} \, \identity
\label{newfreq}
\end{equation}
while the boundary term is
\begin{equation}
{\cal B} = - \frac{1}{2} \int d \tau d^3 k \, \frac{d}{d \tau} \left( \frac{1}{f} \, \frac{d f}{d \tau} \, Y^\dagger Y \right)
\end{equation}

In principle, the two actions~(\ref{azx}) and (\ref{azy}) lead to two different quantization procedures for the fields, and to two different initial conditions for the cosmological application. We want to determine under which conditions these two procedures lead to the same physical result.~\footnote{The physical perturbations will be linear combinations of $X_i$ or, equivalently, $Y_i$. We can equivalently compute the evolution of these physical modes by writing and solving the Einstein equations in terms of  the fields $X_i$ or $Y_i$; the only difference that may arise from using one or the other set of fields is in the initial conditions that the two quantizations (either in terms of the time variable $\eta$ or $\tau$) impose.} Let us now assume that the off-diagonal terms of both $\Omega_\eta$ and $\Omega_\tau$ can be neglected at asymptotically early times (as this is the case for the modes studied in the present work, see subsection~\ref{subinitial}). As discussed after eq.~(\ref{azd}), if the frequency $\Omega_\eta$ is adiabatically evolving, we can set the initial conditions (up to an irrelevant phase)
\begin{equation}
\left( X_i \right)_{\rm in} = \frac{1}{\sqrt{2 \Omega_\eta}} \;\;\;,\;\;\; 
\left( \frac{d X_i}{d \eta} \right)_{\rm in} = - i \, \sqrt{\frac{\Omega_\eta}{2}} \;\;\;,\;\;\; {\rm if} \;\;
\frac{1}{\Omega_\eta^2} \, \frac{d \Omega_\eta}{d \eta} \ll 1
\label{inx}
\end{equation}
where $\Omega_\eta$ here refers to the diagonal entry corresponding to the mode $X_i$.

On the other hand, if $\Omega_\tau$ is also adiabatically evolving, we can also impose
\begin{equation}
\left( Y_i \right)_{\rm in} = \frac{1}{\sqrt{2 \Omega_\tau}} \;\;\;,\;\;\; 
\left( \frac{d Y_i}{d \tau} \right)_{\rm in} = - i \, \sqrt{\frac{\Omega_\tau}{2}} \;\;\;,\;\;\; {\rm if} \;\;
\frac{1}{\Omega_\tau^2} \, \frac{d \Omega_\tau}{d \tau} \ll 1
\label{iny}
\end{equation}
The initial conditions~(\ref{iny}) can be used to set the initial conditions for $X_i$ and $d X_i /  d \eta$ through the relations~(\ref{newtime}) and~(\ref{newfield}). Proceeding in this way, we find
\begin{eqnarray}
\left\{ \left( Y_i \right)_{\rm in} \;,\;\; \left( \frac{d Y_i}{d \tau} \right)_{\rm in} \right\} \;\;\rightarrow \;\;
\left\{ \begin{array}{l}
\left( X_i \right)_{\rm in} = \frac{1}{\sqrt{2 \Omega_\tau} \, f} = \frac{1}{\sqrt{2 \Omega_\eta}} \, \left[ 1 - \frac{f^3}{\Omega_\eta^2} \, \frac{d^2 f}{d \tau^2} \right]^{-1/4} \\ \\
\left( \frac{d X_i}{d \eta} \right)_{\rm in} = - i \sqrt{\frac{\Omega_\eta}{2}} \left[ 1 -  \frac{f^3}{\Omega_\eta^2} \, \frac{d^2 f}{d \tau^2} \right]^{1/4} - \frac{f \, \frac{d f}{d \tau}}{\sqrt{2 \, \Omega_\eta}} \, \left[ 1 -  \frac{f^3}{\Omega_\eta^2} \, \frac{d^2 f}{d \tau^2} \right]^{-1/4}
\end{array} \right.
\end{eqnarray}
If these conditions agree with those obtained by using the fields $X_i$ and the time $\eta$ directly, namely eqs.~(\ref{inx}), then the two quantizations are equivalent. This happens only if
\begin{equation}
f \, \Big\vert \frac{d f}{d \tau} \Big\vert \ll \Omega_\eta \;\;\;,\;\;\; f^3 \, \Big\vert \frac{d^2 f}{d \tau^2} \Big\vert \ll \Omega_\eta^2
\end{equation}
at asymptotically early times. One can verify that these conditions are equivalent to
\begin{equation}
\frac{1}{f} \, \Big\vert \frac{d f}{d \eta} \Big\vert \ll \Omega_\eta \;\;\;,\;\;\; \frac{1}{f} \Big\vert \frac{d^2 f}{d \eta^2} \Big\vert \ll \Omega_\eta^2
\label{conds}
\end{equation}

Let us now discuss under which circumstances these conditions are met. As we mentioned, we can consistently set both (\ref{inx}) and (\ref{iny})  only if both $\Omega_\eta$ and $\Omega_\tau$ are adiabatically evolving (each with respect to its own time). Eq.~(\ref{newfreq}) can be differentiated to give
\begin{eqnarray}
\frac{1}{\Omega_\tau^2} \, \frac{d \Omega_\tau}{d \tau} &=& \frac{1}{2 \, \left[ \Omega_\tau^2 \right]^{3/2}} \, \frac{d \Omega_\tau^2}{d \tau} \nonumber\\
&=&  \frac{ \frac{1}{\Omega_\eta^2} \, \frac{d \Omega_\eta}{d \eta} - \frac{2}{\Omega_\eta} \, \frac{1}{f} \, \frac{d f}{d \eta} + \frac{1}{2 \, \Omega_\eta^3} \left[ - 12 \left( \frac{1}{f} \, \frac{d f}{d \eta} \right)^3 + \frac{9}{f} \, \frac{d f}{d \eta} \, \frac{1}{f} \, \frac{d^2 f}{d \eta^2} - \frac{1}{f} \, \frac{d^3 f}{d \eta^3} \right]}{\left[
1 + \frac{1}{\Omega_\eta^2} \left( \left( \frac{1}{f} \, \frac{d f}{d \eta} \right)^2 - \frac{1}{f} \, \frac{d^2 f}{d \eta^2} \right) \right]^{3/2}}
\label{newadia}
\end{eqnarray}
Barring accidental cancellations between the different terms, both frequencies are adiabatically evolving only if
\begin{equation}
\frac{1}{f} \, \Big\vert \frac{d^\alpha f}{d \eta^\alpha} \Big\vert \ll \Omega_\eta^\alpha \;\;\;,\;\;\; \alpha = 1 ,\, 2 ,\, 3
\label{conds2}
\end{equation}
These conditions also imply that the two quantization procedures are equivalent, cf. eqs.~(\ref{conds}).

To summarize, we can consistently use the two different time variables $\eta$ and $\tau$, provided the relation among them varies sufficiently slowly. If conditions~(\ref{conds2}) are met, and if the frequency $\Omega_\eta$ is adiabatically evolving, then also the frequency $\Omega_\tau$ is adiabatically evolving in its own time. These two procedures lead to equivalent initial conditions for the physical modes. On the other hand, we see that not all choices for conformal time are appropriate for the quantization. 

We proceed by discussing what these findings imply for the Bianchi $I$ backgrounds studied in the main text. For sake of clarity, we study the positive and negative branches separately in the two following sections.

\subsection{Positive branch}
\label{appC1}

We show in Section \ref{subinitial} that, for the positive branch, one can quantize the perturbations
in the conformal time $\eta \,$, defined in eq.~(\ref{back}). Here, we want to investigate whether some other time variable $\tau$ can be used. For definiteness, let us restrict our attention to time redefinitions of the form
\begin{equation}
d \eta = f^{-2} \left( \tau \right) \,  d \tau = a^{-2 \, \alpha} \, b^{-2 \beta} \, d \tau
\label{deftau2}
\end{equation}
where $a$ and $b$ are the two scale factors, and $\alpha$ and $\beta$ two constant parameters (this is the most immediate generalization of what is done in the isotropic case). Eqs.~(\ref{conds2}) give the three conditions that $f$ needs to satisfy for the two choices of time to provide the same initial conditions. We perform the computation using the time $\eta \,$, and we then present the early time expansion in physical time: 
\begin{eqnarray}
\frac{1}{f} \, \Bigg\vert \frac{d f}{d \eta} \Bigg\vert &=& \vert \alpha \vert \, a \left( t \right) H_a \left( t \right) \left[ 1 + {\rm O } \left( m^2 \, t^2 \right) \right] \ll \Omega_\eta \nonumber\\
\frac{1}{f} \, \Bigg\vert \frac{d^2 f}{d \eta^2} \Bigg\vert &=& \alpha^2 \, a^2 \left( t \right) H_a^2 \left( t \right) \left[ 1 + {\rm O } \left( m^2 \, t^2 \right) \right] \ll \Omega_\eta^2 \nonumber\\
\frac{1}{f} \, \Bigg\vert \frac{d^3 f}{d \eta^3} \Bigg\vert &=& \vert \alpha^3  \vert \, a^3 \left( t \right) H_a^3 \left( t \right) \left[ 1 + {\rm O } \left( m^2 \, t^2 \right) \right] \ll \Omega_\eta^3 \nonumber\\
\label{conds3}
\end{eqnarray}
Namely, these three conditions coincide.~\footnote{To be rigorous, we showed that the three conditions~(\ref{conds2}) are sufficient to ensure that the two quantizations are equivalent, but not necessary, since there may be cancellations in eq.~(\ref{newadia}); however, the specific form of (\ref{conds3}) shows that these conditions are also necessary, once we restrict our attention to time redefinitions of the type~(\ref{deftau2}).} From eq.~(\ref{freqearly}) we have $\Omega_\eta^2 = k_1^2 + {\rm O} \left( m^2 \, t^2 \right)$ for all the three modes, where $k_1$ is the longitudinal (comoving) momentum of the mode in the anisotropic $x-$direction. Therefore, we must have $\vert \alpha \vert a \left( t \right) H_a \left( t \right) \ll k_1$ at asymptotically early times. The product $a \, H_a$ is nearly constant all throughout the anisotropic sage, and very close to the value $k_{\rm iso}$ defined in Section \ref{subcurva}. Therefore, the three conditions (\ref{conds3}) become
\begin{equation}
k_1 \gg \vert \alpha \vert \, k_{\rm iso}
\end{equation}

The modes with $k_1 \gg k_{\rm iso}$ are those leaving the horizon after the universe has become isotropic, and we recover the standard result for them (see Section \ref{subcurva}). The modes with smaller longitudinal momentum are those sensitive to the evolution of the universe during the anisotropic stage, and therefore are those of real interest for the present analysis. We can only quantize them using a time variable for which $\alpha = 0$ in eq.~(\ref{deftau2}). We remark that $b$ is constant at asymptotically early times. Therefore, if we use the scale factors in the definition of the conformal time, the time $\eta$ used in the main text is the only possible one, up to a trivial constant rescaling.

\subsection{Negative branch}
\label{appC2}

Let us start from the early time frequencies for the modes canonically normalized with respect to conformal time $\eta$. As shown in eq.~(\ref{freqearly2}) we find
\begin{eqnarray}
\Omega_\eta^2 &=& a^2 \left[ - \frac{5}{9 t^2} + \frac{k_2^2}{b^2} + {\rm O} \left( 1 \right) \right] \quad\quad,\quad\quad 2 \, {\rm d \; vector} \nonumber\\
\Omega_\eta^2 &=& a^2 \left[ \frac{4}{9 t^2} + \frac{k_2^2}{b^2} + {\rm O} \left( 1 \right) \right] \quad\quad,\quad\quad 2 \, {\rm d \; scalar} 
\end{eqnarray}
where, in the $2$d scalar case, $\Omega_\eta^2$ refers to the dominant diagonal entries of the frequency matrix (which are identical). We recall that $a \propto t^{-1/3} ,\; b \propto t^{2/3}$ at asymptotically early times (as always, although the frequency refers to the conformal time $\eta$, we show the expansion in terms of physical time $t$). As discussed in the main text, these frequencies are not adiabatically evolving at early times; moreover the $2$d vector mode is tachyonic. Let us discuss whether the situation improves if we use a different time variable $\tau$; as for the positive branch, we consider time redefinitions of the form (\ref{deftau2}). 

Let us discuss the $2$d vector mode first. Evaluating the two equations~(\ref{newfreq}) and ~(\ref{newadia}) we find
\begin{eqnarray}
\Omega_\tau^2 &=& \frac{a^{-4 \alpha + 2} \, b^{-4 \beta}}{9 \, t^2} \left[ \left( \alpha - 2 \beta + 1 \right) \left( \alpha - 2 \beta - 5 \right) \right] \left[ 1 + {\rm O} \left( m^{2/3} t^{2/3} \right) \right] \nonumber\\
\frac{1}{\Omega_\tau^2} \, \frac{d \Omega_\tau}{d \tau} &=& \frac{ \left[ \alpha - 2 \left( 1 + \beta \right) \right]}{\sqrt{\left( \alpha - 2 \beta + 1 \right) \left( \alpha - 2 \beta - 5 \right)}} \left[ 1 + {\rm O} \left( m^{2/3} t^{2/3} \right) \right]
\end{eqnarray}
We can obtain an adiabatic initial evolution of the frequency if we choose the parameters of the transformation such that $\alpha = 2 \left( 1 + \beta \right)$. However, in this case, the frequency $\Omega_\tau$ becomes
\begin{equation}
\Omega_\tau^2 = - \frac{a^{-8 \beta - 6} b^{-4 \beta}}{t^2} \left[ 1 + {\rm O} \left( m^{2/3} t^{2/3} \right) \right]
\label{omtvm}
\end{equation}
By using the early time asymptotics for the scale factors, we can see that the frequency
(\ref{omtvm}) approaches a constant negative value close to the initial singularity. This leads to a tachyonic growth of the mode, and to the breakdown of the linearized computation (see the main text).

Let us finally discuss the $2$d scalar sector. For generic values of $\alpha$ and $\beta$ we now find
\begin{eqnarray}
\Omega_\tau^2 &=& \frac{a^{-4 \alpha + 2} \, b^{-4 \beta}}{9 \, t^2} 
\left\{ \left[ \alpha - 2 \left( 1 + \beta \right) \right]^2 + 9 \, t^2 \, k_2^2 / b^2 \right\} + {\rm O} \left( 1 \right) \nonumber\\
\frac{1}{\Omega_\tau^2} \, \frac{d \Omega_\tau}{d \tau} &=& \frac{ 2 \left[ \alpha - 2 \left( 1 + \beta \right) \right]^3 + 9 \, t^2 \left( - 3 + 2 \alpha - 4 \beta \right) k_2^2 / b^2 + {\rm O} \left( t^2 \right)}{\left[ \left[ \alpha - 2 \left( 1 + \beta \right) \right]^2 + 9 \, t^2 \, k_2^2 / b^2 + {\rm O} \left( t^2 \right) \right]^{3/2}}
\end{eqnarray}
For $\alpha \neq 2 \left( 1 + \beta \right)$, the second expression evaluates to $2$ at asymptotically early time. For $\alpha = 2 \left( 1 + \beta \right)$, the dominant term vanishes both at the numerator and at the denominator, and the adiabaticity condition approaches $1 / \left( 3 \, t \, p_2 \right)$. This quantity diverges as $t^{-1/3}$ as the time approaches the initial singularity at $t = 0 \,$. Therefore, no time variable $\tau$ chosen as in~(\ref{deftau2}) leads to an initial adiabatic evolution for the $2$d scalars.

\end{document}